\documentclass[lettersize,journal]{IEEEtran}
\usepackage{amsmath,amsfonts}
\usepackage{algorithmic}
\usepackage{array}
\usepackage{textcomp}
\usepackage{stfloats}
\usepackage{url}
\usepackage{verbatim}
\usepackage{color}
\usepackage{graphicx} 
\usepackage{subfigure}
\usepackage{cite}
\usepackage{xcolor}
\usepackage{tcolorbox}
\usepackage{color}
\usepackage{svg}

\usepackage[linesnumbered,ruled,vlined]{algorithm2e}%[ruled,vlined]{  
\def\BibTeX{{\rm B\kern-.05em{\sc i\kern-.025em b}\kern-.08em 
    T\kern-.1667em\lower.7ex\hbox{E}\kern-.125emX}}
\usepackage{balance}

\begin{document} 

\title{WiCAL: Accurate Wi-Fi-Based 3D Localization Enabled by Collaborative Antenna Arrays \\ 
\thanks{This work is supported by the National Natural Science Foundation of China under Grant 12141107, the Key Research and Development Program of Wuhan under Grant 2024050702030100, and the Interdisciplinary Research Program of HUST (2023JCYJ012). ({\it Corresponding author: Tiebin Mi})}
\thanks{F. Wang is with the School of Electronic Information and Communications, Huazhong University of Science and Technology, Wuhan 430074, China, and also with the Institute of Artificial Intelligence, Huazhong University of Science and Technology, Wuhan 430074, China. (e-mail: wangfuhai@hust.edu.cn)}
\thanks{Z. Li, R. Xiong, T. Mi, and R. Qiu are with the School of Electronic Information and Communications, Huazhong University of Science and Technology, Wuhan 430074, China (e-mail: lizhe22@hust.edu.cn; rujing@hust.edu.cn; mitiebin@hust.edu.cn; caiming@hust.edu.cn).}
}

\author{\IEEEauthorblockN{Fuhai Wang,~\IEEEmembership{Student Member,~IEEE,} 
Zhe Li, 
Rujing Xiong,~\IEEEmembership{Student Member,~IEEE,} \\
Tiebin Mi,~\IEEEmembership{Member,~IEEE,} 
and Robert Caiming Qiu,~\IEEEmembership{Fellow,~IEEE}} \\
}

\maketitle
\thispagestyle{empty}
\begin{abstract} 
    Accurate 3D localization is essential for realizing advanced sensing functionalities in next-generation Wi-Fi communication systems. This study investigates the potential of multistatic localization in Wi-Fi networks through the deployment of multiple cooperative antenna arrays. The collaborative gain offered by these arrays is twofold: (i) intra-array coherent gain at the wavelength scale among antenna elements, and (ii) inter-array cooperative gain across arrays. To evaluate the feasibility and performance of this approach, we develop WiCAL (Wi-Fi Collaborative Antenna Localization), a system built upon commercial Wi-Fi infrastructure equipped with uniform rectangular arrays (URAs). These arrays are driven by multiplexing embedded radio frequency (RF) chains available in standard access points or user devices, thereby eliminating the need for sophisticated, costly, and power-hungry multi-transceiver modules typically required in multiple-input and multiple-output (MIMO) systems. To address phase offsets introduced by RF chain multiplexing, we propose a three-stage, fine-grained phase alignment scheme to synchronize signals across antenna elements within each array. A bidirectional spatial smoothing MUSIC algorithm is employed to estimate angles of arrival (AoAs) and mitigate performance degradation caused by correlated interference. To further exploit inter-array cooperative gain, we elaborate on the synchronization mechanism among distributed URAs, which enables direct position determination by bypassing intermediate angle estimation. Once synchronized, the distributed URAs effectively form a virtual large-scale array, significantly enhancing spatial resolution and localization accuracy. WiCAL is validated using $3 \times 4$ URAs operating at the 5.2 GHz band. Experimental results demonstrate median AoA estimation errors of $1^\circ$ in elevation and $1.5^\circ$ in azimuth under intra-array coherent processing. For inter-array collaboration, the system achieves a median localization error of 15.6 cm using two URAs, outperforming state-of-the-art methods.
\end{abstract}


\section{Introduction}
\IEEEPARstart{A}{ccurate} 3D localization is a fundamental feature of Wi-Fi sensing, especially in future integrated sensing and communication systems. Emerging applications demand low-cost, high-accuracy positioning solutions, including indoor positioning~\cite{zhang2019efficient,10592075}, low-altitude economy~\cite{geraci2022integrating,10608169}, autonomous navigation~\cite{ayyalasomayajula2020deep,arun2024wais}, wireless communications~\cite{tedeschini2024real,xu2024mimo} and virtual reality~\cite{kotaru2017position,10599121}, among others. Antenna array techniques, as a key enabler of wireless communications, are now increasingly adopted in Wi-Fi networks. In this context, mimicking multiple-input and multiple-output (MIMO) techniques from cellular networks represents an initial attempt \cite{xiong2013arraytrack}. Nevertheless, unlike cellular infrastructure, both access points (APs) and user devices in Wi-Fi networks are highly cost-sensitive~\cite{wang2013spectral}, necessitating a careful trade-off between array performance and implementation complexity. 

% Power splitter-based methods configure a dedicated antenna to achieve synchronization among antennas distributed across multiple Wi-Fi network interface cards (NICs). However, in addition to low synchronization efficiency, a major limitation is that channel state information (CSI) acquisition must be conducted separately at individual APs, thus precluding efficient utilization of a unified command instruction across multiple APs.

A cost-effective approach in Wi-Fi systems utilizes a limited number of radio frequency (RF) chains to support a larger antenna array, thereby eliminating the need for a dedicated RF chain per antenna element. Current antenna array extension methods can be broadly categorized into two classes: power splitter-based methods~\cite{gjengset2014phaser} and RF switch-based methods~\cite{xie2018swan}. Power splitter-based methods configure a dedicated RF chain to achieve synchronization among antennas distributed across multiple Wi-Fi network interface cards (NICs). An inherent limitation of this approach is its low synchronization efficiency. Furthermore, despite achieving antenna-level alignment, a notable drawback is the asynchronous CSI acquisition across NICs, which necessitates distributed processing of subsequent algorithms and significantly increases overall system complexity. Switch-based methods represent a class of multiplexing techniques favored for their efficient utilization of RF resources in Wi-Fi networks~\cite{gu2021tyrloc}. Notable examples include iArk~\cite{an2020general} and SWAN~\cite{xie2018swan}. Subsequent studies have demonstrated that such methods enable the extraction of more precise parameters including angle of arrival (AoA)~\cite{kotaru2015spotfi,he2020multi}, time of flight~\cite{song2022rf,soltanaghaei2018multipath}, and Doppler frequency shifting~\cite{qian2018widar2,chen2023cross}. These parameters have become key indicators for determining the position of devices or users in various environments.

From the perspective of antenna arrays, the collaborative gain achieved through multiple arrays is twofold. First, intra-array coherent gain at the wavelength scale---enabled by synchronized signals among antenna elements within a single array---facilitates high-resolution AoA estimation. Second, inter-array collaboration across arrays deployed at different stations further enhances positioning accuracy. In RF chain multiplexing schemes, a primary challenge in achieving wavelength-scale coherent gain lies in compensating for phase offsets introduced during the multiplexing process. Antenna synchronization is particularly difficult in Wi-Fi systems due to hardware heterogeneity. Even when synchronization is achieved, mitigating the effects of correlated signal interference remains essential, as such interference is common in practical Wi-Fi environments due to multipath propagation. On the other hand, obtaining cooperative gain at the inter-array level requires elaborate algorithmic design. The core challenge lies in estimating the true position from noisy and imperfect measurements collected across distributed APs. While several studies have explored Wi-Fi-based 2D localization~\cite{sesyuk2022survey,10592075}, extending these approaches to robust 3D localization remains an open and challenging problem.

\subsection{Contributions}

This paper presents Wi-Fi Collaborative Antenna Localization (WiCAL), a novel framework for high-precision 3D localization in Wi-Fi networks. At the hardware level, WiCAL adopts an innovative architecture based on switched antenna arrays, enabling a device with a limited number of RF chains to efficiently drive larger antenna arrays. When combined with advanced data fusion algorithms, this architecture supports joint localization across multiple uniform rectangular arrays (URAs). The main contributions of this work are summarized as follows:

\begin{itemize} 
\item \textbf{Collaborative Antenna Array Design.} We present WiCAL, a practical framework for collaborative antenna arrays designed for 3D localization in Wi-Fi networks. These arrays are driven by multiplexing embedded RF chains available in access points or user devices. Unlike the sophisticated multi-transceiver modules typically employed in conventional MIMO systems, the switch-based design significantly reduces both hardware cost and system complexity. To address the phase offsets introduced by RF chain multiplexing, a three-stage, fine-grained phase alignment method is developed to synchronize the CSI across antenna elements. 

\item \textbf{3D Localization Algorithms.} We propose a two-step algorithm for multistatic 3D localization. The first step leverages synchronized signals among antenna elements and employs a bidirectional spatial smoothing MUSIC algorithm to estimate AoAs while mitigating performance degradation caused by correlated interference due to multipath propagation. The second step exploits inter-array cooperative gain by applying a robust closest-point estimation method to approximate the source location from the set of AoA estimates. In addition, we develop an inter-array synchronization framework that enables direct position determination (DPD) by bypassing intermediate AoA estimation. Once synchronized, the distributed URAs effectively constitute a virtual large-scale array, substantially enhancing spatial resolution and localization accuracy. Finally, a progressive local traversal strategy is introduced to iteratively refine the position estimate.

\item \textbf{Performance Validation.} We develop a hardware prototype to evaluate the performance of WiCAL. Extensive testing is conducted across diverse scenarios, including multi-source 3D AoA estimation, localization, and tracking. Experimental results demonstrate median AoA errors of $1^\circ$ in elevation and $1.5^\circ$ in azimuth under intra-array coherent processing. For inter-array collaboration, WiCAL achieves a median localization error of 15.6 cm using two URAs. To the best of our knowledge, this represents the first system to achieve state-of-the-art 3D localization using only commercial Wi-Fi devices.
\end{itemize}

\subsection{Prior Works}
\subsubsection{3D Localization} 
Wireless-based positioning technologies broadly include Bluetooth, UWB, mmWave, and Wi-Fi, each offering distinct advantages tailored to specific application scenarios. Bluetooth is a low-cost solution; however, its dependence on received signal strength limits its accuracy to a few meters, making it unsuitable for high-precision applications~\cite{farahsari2022survey}. UWB, a short-range technology utilizing wide bandwidths, can achieve centimeter-level accuracy through time of arrival measurements. Nevertheless, it suffers from limited range and is highly susceptible to signal attenuation due to obstructions~\cite{farahsari2022survey}. Although mmWave technology holds promise for high-precision 3D localization, it continues to face challenges such as high power consumption and sensitivity to environmental conditions~\cite{sesyuk2022survey}.

Wi-Fi, widely used in indoor environments, supports localization through signal strength~\cite{liu2013accurate,xiong2024fair}, CSI~\cite{tai2019toward,zhao2023nerf2}, or time of flight measurements~\cite{zhang20193d,qian2017enabling,wu2021witraj,chen2023cross}. Phase-synchronized CSI data enables the application of AoA-based localization methods. Existing AoA estimation algorithms encompass a diverse array of techniques, including MUSIC, ESPRIT, Capon, SPICE and compressive sensing (CS)-based approaches~\cite{stoica2005spectral,zhang20193d}. However, challenges remain due to correlated signals, the limited number of antennas, and the increased computational complexity introduced by the higher dimensionality of 3D localization tasks.

\subsubsection{Distributed Array Localization} 
The distributed array offers a range of benefits, including greater spatial diversity and improved reliability~\cite{nanzer2021distributed}. AoA-based methods are known for their high accuracy but require effective synchronization and aperture synthesis of distributed antenna arrays~\cite{nanzer2021distributed}. A joint 2D AoA estimation using distributed Wi-Fi is highlighted in~\cite{yang2023multiple}. This study proposes algorithmic innovations that outperform previous works like SpotFi~\cite{kotaru2015spotfi}. However, coherent distributed localization presents a significantly harder challenge, as it requires aligning the relative electrical states of the antennas in situ. This entails ensuring that all nodes operate at the same frequency, calibrating phase errors from internal subsystem delays, and estimating and correcting phase and timing differences caused by the nodes' relative locations. To tackle emerging 3D distributed localization challenges, our study introduces phase synchronization and data-level fusion computation among distributed arrays as potential solutions.

\begin{figure} 
    \centerline{\includegraphics[width=0.95\columnwidth]{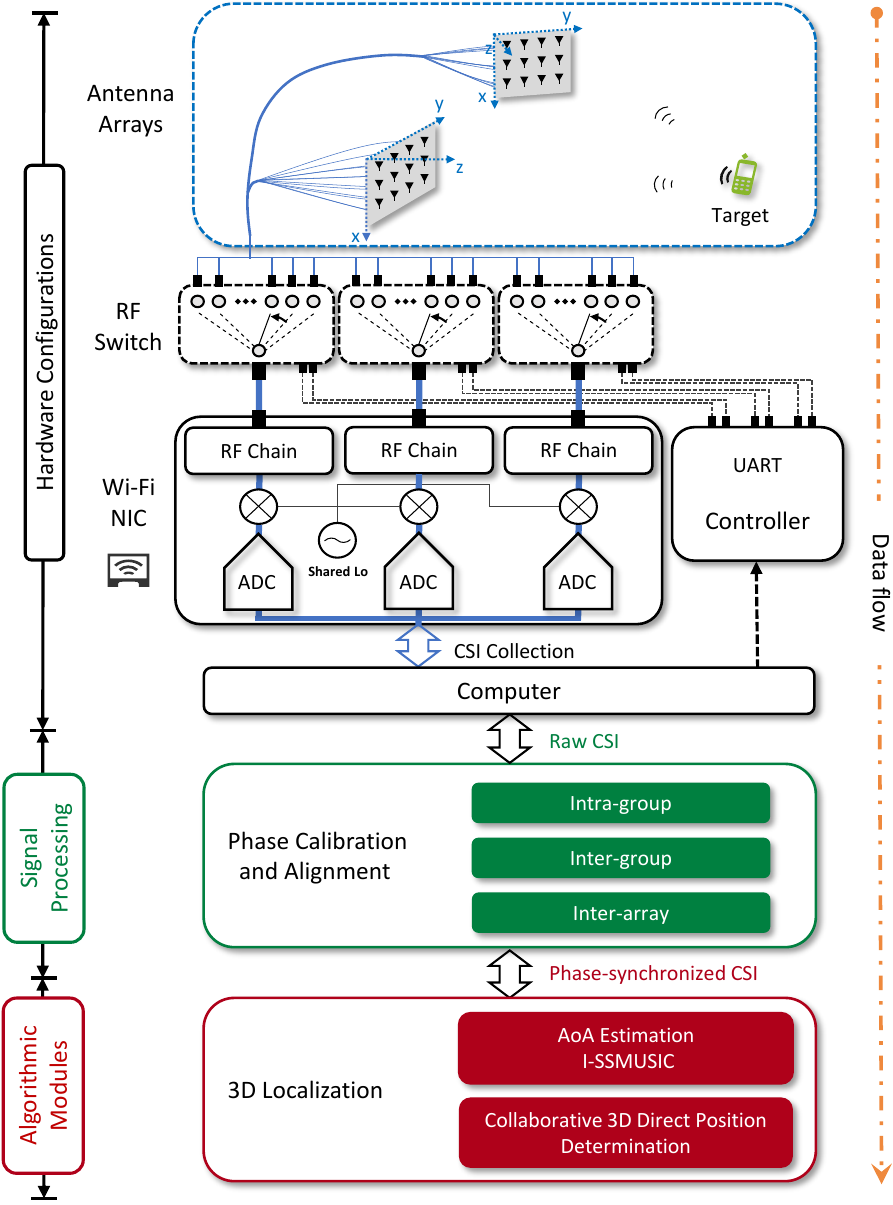}}  
    \caption{The system overview of WiCAL.} 
    \label{diagram}  
\end{figure}

\subsubsection{Switch-based Antenna Array} 
Compared to phase shifters used in conventional MIMO and numerous devices, switches-based architectures reduce cost, complexity, and power consumption. IArk~\cite{an2020general} is a switches-based architecture and utilizes a single 1.5 m $\times$ 1.5 m antenna array for 3D localization, coupled with specialized USRP equipment. Unlike our approach, iArk requires an additional RF port for phase calibration to achieve phase alignment across all antennas. Moreover, relying on a single antenna board restricts its spatial degrees of freedom. Additionally, iArk employs a supervised AI method for signal fusion. In contrast, SWAN~\cite{xie2018swan} is a Wi-Fi-based localization scheme, but it is restricted to 2D localization. SWAN does not further consider the multi-angle distributed localization and focuses solely on angle estimation. Beyond this, previous studies~\cite{xie2018swan,an2020general,xiong2013arraytrack,gjengset2014phaser} rely on multiple synchronous RF chains, with at least one chain always connected to a reference antenna for CFO calibration~\cite{gu2021tyrloc}. In contrast, we propose a high-utilization RF chain approach. Using a commercial Wi-Fi device with three synchronous RF chains, a three single-pole-N-throws (SPNT) switch expansion scheme can extend the array to 3N antennas, offering high multiplexing capability. Previous setups, where a RF chain is fixed to a reference antenna, support a maximum of 2N+1 antennas.

The remainder of this paper is organized as follows. Section~\ref{Design of Switched Antenna Array} details the design of switched antenna array and phase calibration. A geometric-based 3D fast positioning algorithm is introduced in Section~\ref{Geometric-based Positioning}, while the proposed collaborative 3D direct position determination is described in Section~\ref{Joint 3D Direct Position Determination}. Section~\ref{Prototype} outlines the prototype implementation and experimental environments of the proposed WiCAL. Performance evaluation is presented in Section~\ref{Evaluation}, followed by conclusion and future work in Section~\ref{Conclusion}.

\section{Switched Antenna Array and Phase Calibration}\label{Design of Switched Antenna Array} 
The architecture of WiCAL comprises three main components: a hardware configuration module, a signal processing module, and a 3D localization algorithmic module, as illustrated in Fig.~\ref{diagram}. The hardware consists of a switched URA, which enables the multiplexing of a limited number of RF chains across the antenna elements. This design eliminates the need for additional phase shifters, thereby avoiding the high hardware costs typically associated with integrating new components.

\begin{figure} 
    \centerline{\includegraphics[width=0.65\columnwidth]{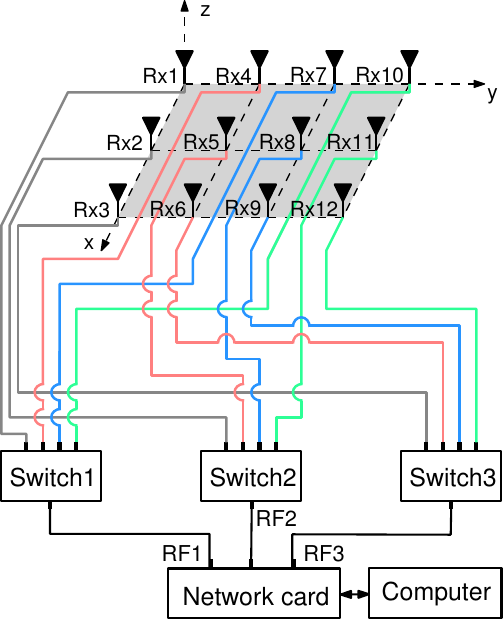}}  
    \caption{Switched URA. Each port of the RF switches is connected to a unique antenna, with every trio of antennas grouped together to capture synchronized CSI data through Wi-Fi device. Each group is marked with the same color. For instance, by employing three SP4T RF switches, we can collect CSI data from a total of 12 antennas, organized into 4 distinct groups.} 
    \label{antenna configuration}  
\end{figure}

\subsection{Multiplexing Protocol and Grouping} \label{Switched Antenna Expansion and Grouping}

To support the antenna multiplexing with a limited number of RF chains, we develop a connection protocol designed to maximize RF chain usage. Given a standard commercial Wi-Fi NIC equipped with three synchronized RF chains, a straightforward expansion scheme employing three single-pole-N-throw (SPNT) RF switches enables the array to scale up to 3N antenna elements. In prior synchronization approaches~\cite{xie2018swan,an2020general,xiong2013arraytrack}, one dedicated RF chain in each NIC is permanently connected to a reference antenna, thereby limiting the system to a maximum of 2N+1 antennas.

The core idea of our protocol is that each port of RF switches is connected to a distinct antenna. Every set of three antennas is grouped for subsequent alignment processing. For example, by employing three SP4T switches, we can collect CSI data from twelve antenna elements, organized into four distinct groups, as illustrated in Fig.~\ref{antenna configuration}. The antennas in each group, specifically $\{1,2,3\}$, $\{4,5,6\}$, $\{7,8,9\}$, and $\{10,11,12\}$, are connected to $\text{RF}_1$, $\text{RF}_2$ and $\text{RF}_3$, respectively, through the RF switches.

\subsection{Phase Calibration and Alignment} \label{Phase Calibration}

Before delving into the specific phase calibration and alignment algorithms, we first analyze the key factors contributing to phase offsets. 
In general, phase offsets can be categorized into three types: {\it intra-group} unknown phase offsets $\xi_1$, which arise from the initial phase discrepancies of phase-locked-loop (PLL) among RF chains and external paths (e.g., cables and RF switches)~\cite{xiong2013arraytrack}; {\it inter-group} random phase offsets $\xi_2$, which are primarily introduced by carrier frequency offset (CFO)~\cite{xie2018swan}; and {\it inter-array} phase offsets across multiple URAs, which result from distance differences between the signal source and each URA.

\subsubsection{Intra-group phase calibration} Theoretically, the received CSI at frequency $f$ for an antenna array can be expressed as
\begin{equation} \label{received signals} 
    CSI(t,m,f) = e^{-j 2 \pi f_{\mathrm{CFO}} t } \times \alpha e^{j \varphi_{m}} e^{-j 2 \pi f \left(\tau_m^c+\tau_m^p\right)},
\end{equation} 
where $t$ and $m$ denote the index of time and antenna, respectively, $f_{\mathrm{CFO}}$ represents the CFO, $\alpha$ denotes the attenuation coefficient of the signal, and $\varphi$ denotes the PLL initial phase. The external path delay is composed of the external cable delay $\tau^c$ and the propagation path delay $\tau^p$. Note that the CFO is consistent across different RF chains and therefore does not introduce phase offsets among the three antennas within a group. In array-based AoA estimation, the phase difference due to the time of flight difference $\tau_{1,m}^p$  between the $m$-th antenna and the first antenna plays a critical role. Consequently, the unknown phase offsets caused by the PLL initial phase $\varphi$ and external cable delays $\tau^c$~\cite{xiong2013arraytrack} within a synchronous group must be properly addressed.

\begin{figure}
    \centerline{\includegraphics[width=1\columnwidth]{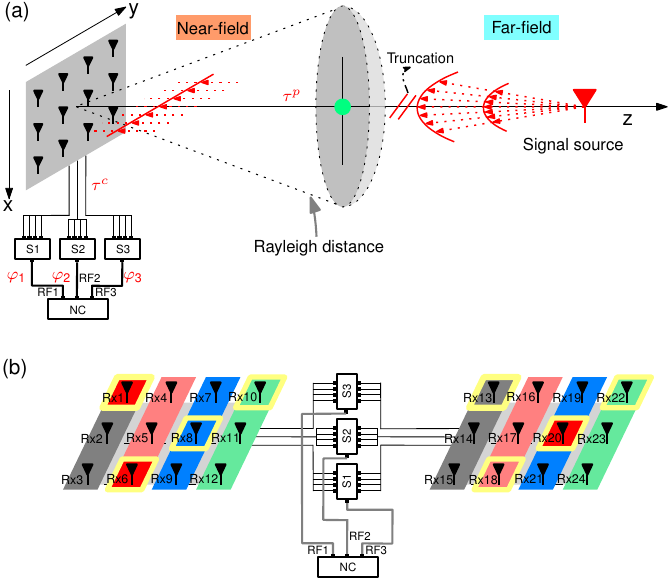}} 
    \caption{(a) The signal source is placed in front of the antenna array, within the far-field, ensuring that $\tau^{p}$ remains consistent across all antennas. (b) The CPAs for inter-group calibration includes antennas \{1, 6, 8, 10\} and \{13, 18, 20, 22\}, enclosed in yellow boxes. Antennas \{1, 6, 20\}, marked in red, are used for phase alignment between arrays.} 
    \label{farfield} 
\end{figure} 

We present a user-friendly method for calibrating phase offsets, as illustrated in Fig.~\ref{farfield}(a). Specifically, the signal source is placed directly in front of the antenna array and within the far-field region to ensure that $\tau^{p}$ remains consistent across all antennas (i.e., $\tau^{p}_1 = \cdots = \tau^{p}_M$). Under these conditions, the observed phase differences are primarily attributed to offsets introduced by the PLL and external hardware components, such as cables and RF switches. For each group, we detail the measured phase offsets between the first antenna and the other two antennas as 
\begin{equation} \label{offsets12}
\begin{aligned}
\hat{\varphi}_{12} =& \varphi_2 + 2 \pi f({\tau}_{2}^{c}+{\tau}_{2}^{p} )- [\varphi_1 + 2 \pi f({\tau}_{1}^{c}+{\tau}_{1}^{p})] \\
                   =& \varphi_{12} +  2 \pi f\tau_{12}^{c},
\end{aligned}
\end{equation}
where $\varphi_{12} = \varphi_2 - \varphi_1$ and $\tau_{12}^{c} = \tau_2^{c} - \tau_1^{c}$. Similarly, $\hat{\varphi}_{13} =\varphi_{13} +  2 \pi f\tau_{13}^{c}$, where $\varphi_{13} = \varphi_3 - \varphi_1$, $\tau_{13}^{c} = \tau_3^{c} - \tau_1^{c}$ and ${\tau}_{1}^{p} = {\tau}_{3}^{p}$.

By subtracting the measured phase offsets from the incoming signals, as shown in Fig.~\ref{phase calibration}(a), the unknown phase differences are effectively canceled out, yielding the corrected intra-group CSI phase offsets $\hat{\varphi}_{g,12}$ and $\hat{\varphi}_{g,13}$ for the $g$-th group.

\subsubsection{Inter-group phase alignment} The CSI measurements for different groups are collected in a time-division manner, resulting in phase misalignment between groups. The key to aligning CSI across groups lies in estimating the inter-group random phase offsets, which arise from the time-varying phase term $2 \pi f_{\mathrm{CFO}} t$. We propose an efficient method that leverages the flexibility of RF switch selection to enable redundant CSI sampling on a dedicated antenna within each group, thereby achieving higher RF chain utilization compared to approaches based on redundant antennas~\cite{xie2018swan}. These dedicated stand-alone antennas, referred to as calibration point antennas (CPAs), facilitate accurate phase alignment across data packets.

\begin{figure}
    \centerline{\includegraphics[width=0.78\columnwidth]{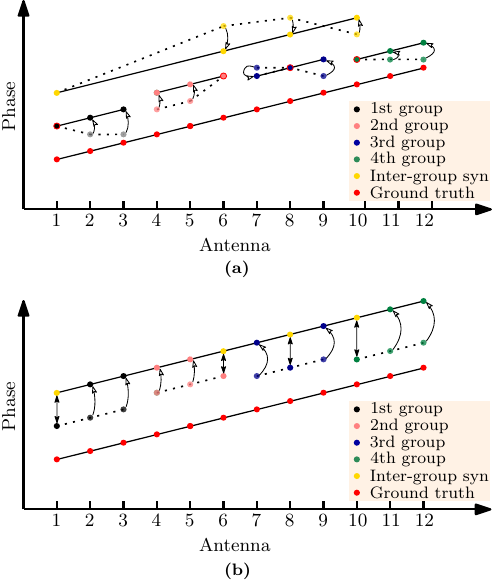}} 
    \caption{Phase calibration and alignment. (a) The intra-group calibration eliminates the phase offsets introduced by radio chains. (b) To synthesize a virtually phase-synchronized CSI from individual CSI measurements, we use the inter-group synchronization to align the phase shifts caused by CFO.} 
    \label{phase calibration}
\end{figure}

For instance, we partition the 12 antennas into four groups, as outlined in Section~\ref{Switched Antenna Expansion and Grouping}. The antennas \{1, 6, 8, 10\} are designated as CPAs, and redundant samplings are conducted on the two other antenna groups \{1, 6, 8\} and \{6, 8, 10\}, respectively. First, the initial phase errors within all groups are eliminated using the intra-group phase calibration method. Then, phase alignment is carried out on antennas \{1, 6, 8, 10\} using the redundant CSI measurements. Finally, the phases of the CPAs within each group are aligned to the reference phases obtained from inter-group synchronization, as illustrated in Fig.~\ref{phase calibration}(b). This process ultimately enables aperture synthesis for the antenna array expanded via RF switches, allowing the aligned CSI to be reliably leveraged for advanced signal analysis.

\subsubsection{Inter-array phase alignment} The deployment of multiple URAs is essential for 3D localization, as it enables diverse spatial perspectives through distributed measurements. Moreover, phase synchronization among these distributed arrays can be implemented using RF switches. By controlling the RF switches, we utilize three RF chains to measure the phase difference between two distributed arrays. Let the positions of the two URAs be denoted by $\boldsymbol{c}_1=(x_1,y_1,z_1)$ and $\boldsymbol{c}_2=(x_2,y_2,z_2)$, respectively. For a signal source located at $\boldsymbol{c}=(x,y,z)$, the phase difference between the signals received by the two URAs can be expressed as
\begin{equation} \label{gamma}
    \bigtriangleup \gamma (\boldsymbol{c}) = 2 \pi \left(\left\|\boldsymbol{c}-\boldsymbol{c}_1 \right\|_2 - \left\| \boldsymbol{c}-\boldsymbol{c}_2 \right\|_2\right) / \lambda ,
\end{equation} 
where $\lambda$ denotes the wavelength and $\left\| \cdot \right\|_2$ represents the Euclidean norm. The phase differences among antenna elements are essential for estimating the location of the signal source. 

Using three SP8T switches, up to 24 antennas can be multiplexed, as illustrated in Fig.~\ref{farfield}(b). The CSI of antennas \{1, 6, 20\}, highlighted in red, is captured to measure the inter-array phase difference $\bigtriangleup \hat{\gamma}$ between the two URAs. An advantage of this multiplexing scheme is its full compatibility with existing Wi-Fi communication protocols, as it operates entirely within the physical layer and requires only minimal control at the data link layer, ensuring seamless integration with current and future Wi-Fi standards.

\subsection{URA Antenna Spacing Design} \label{URA Design}  
The relationship between the scannable region in the sensing space and the maximum antenna element spacing $d_{\max}$ is given by~\cite{edition2009antennas}
\begin{equation}
d_{\max }=\frac{\lambda}{1+ \left| \sin \theta_L \right| },\\ 
\end{equation}
where  $\theta_L$ represents the largest angle to be scanned away from the broadside. In practical applications, due to aperture-impedance mismatch and associated gain degradation, planar antennas are rarely designed to scan more than $120^\circ$ in a single plane~\cite{stutzman2012antenna}. To ensure robust performance, we design a URA with a maximum scan angle of $60^\circ$ and an element spacing of $d = 0.54 \lambda$. Compared to arrays with a half-wavelength spacing, the increased antenna spacing can potentially enhance angular resolution owing to the larger antenna aperture~\cite{van2002optimum}.

\section{A Geometric-based 3D Fast Positioning} \label{Geometric-based Positioning}

A two-step algorithm is developed for multistatic 3D localization. In the first step, an improved MUSIC algorithm is employed to exploit intra-array coherent gain at the wavelength scale while mitigating degradation from correlated interference. The second step leverages inter-array cooperative gain across multiple arrays by applying a robust closest-point estimation algorithm to approximate the true intersection point from a set of AoA estimates.

\subsection{Signal Modeling for URA Antenna}
\begin{figure} 
    \centerline{\includegraphics[width=0.6\columnwidth]{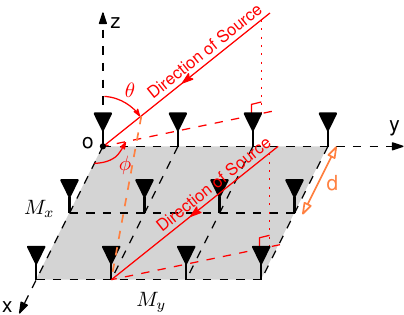}} 
    \caption{URA consisting of $M_x \times M_y$ antennas, where $M_x$ and $M_y$ denote the number of antennas along the x and y axes.} 
    \label{URA_figure}  
\end{figure}

A comprehensive understanding of the forward measurement process associated with the URA is vital for achieving high-resolution AoA estimation. As shown in Fig.~\ref{URA_figure}, a URA comprises $M = M_x \times M_y$ antennas, uniformly spaced by $d$ in both the $X$ and $Y$ direction on the $xoy$-plane, and a spherical coordinate system is employed to represent the direction of the incident plane waves. Let $L$ denote the number of incident sources. For the $l$-th source, the elevation angle $\theta \in [0,\pi/2]$ is measured downward from the $z$ axis, and azimuth angle $\phi \in [0,2\pi)$ is measured counterclockwise from the $x$ axis. We introduce the steering vector $\boldsymbol{a}\left(\theta_{l}, \phi_{l}\right)$ to characterize the additional phase encountered by each antenna relative to the first one. For a URA, the steering vector can be expressed as
\begin{equation} \label{a_theta_phi}
    \boldsymbol{a}\left(\theta_{l}, \phi_{l}\right)
    = \boldsymbol{a}_y\left(\theta_{l}, \phi_{l}\right) \otimes \boldsymbol{a}_x\left(\theta_{l}, \phi_{l}\right),
\end{equation}
where $\otimes$ denotes the kronecker product. The vectors $\boldsymbol{a}_x\left(\theta_{l}, \phi_{l}\right)$ and $\boldsymbol{a}_y\left(\theta_{l}, \phi_{l}\right)$ represent the steering vectors along the $X$- and $Y$- directions, respectively, and are given by
\begin{equation} \label{ax}
    \boldsymbol{a}_{x}(\theta, \phi)
=\left[1 \quad u(\theta, \phi) \quad \cdots \quad u(\theta, \phi)^{M_x-1}\right]^{\top},
\end{equation}
\begin{equation} \label{ay}
    \boldsymbol{a}_{y}(\theta, \phi)
=\left[1 \quad v(\theta, \phi) \quad \cdots \quad v(\theta, \phi)^{M_y-1}\right]^{\top},
\end{equation}
where $u(\theta, \phi)=e^{j 2 \pi d/ \lambda \sin \theta \cos \phi }$ and $v(\theta, \phi)=e^{j 2 \pi d / \lambda \sin \theta \sin \phi}$.

The forward measurement process of a URA can be expressed in matrix form as
\begin{equation} \label{yt}
    \boldsymbol{y}(t)=\boldsymbol{A}(\theta,\phi) \cdot \boldsymbol{s}(t) + \boldsymbol{\epsilon}(t) ,
\end{equation}
where $\boldsymbol{y}(t) \in \mathbb{C}^{M \times 1}$ represents the received signal vector across the $M$ antennas at time $t$. The matrix $\boldsymbol{A}=\left[ \boldsymbol{a}(\theta_1,\phi_1) \ \boldsymbol{a}(\theta_2,\phi_2) \ \cdots \ \boldsymbol{a}(\theta_L,\phi_L) \right]$ denotes the steering matrix, where each column corresponds to a distinct direction of arrival. The vector $\boldsymbol{s}(t)\in \mathbb{C}^{L \times 1}$ contains the source signals, and $\boldsymbol{\epsilon}(t)$ is a complex Gaussian noise vector with zero mean and variance $\sigma^2$. 

In the multi-snapshot scenario, the forward $T$-measurement process is described as
\begin{equation} \label{YT}
    \boldsymbol{Y} = \boldsymbol{A} \cdot \boldsymbol{S} + \boldsymbol{N},
\end{equation}
where $\boldsymbol{Y}=[\boldsymbol{y}(1),\cdots,\boldsymbol{y}(T)]\in \mathbb{C}^{M \times T}$ is the matrix of received signals across $T$ time snapshots, $\boldsymbol{S}=[\boldsymbol{s}_1,\cdots,\boldsymbol{s}_T]\in \mathbb{C}^{L \times T}$ denotes the source signal matrix, and $\boldsymbol{N}\in \mathbb{C}^{M \times T}$ represents the noise.

\subsection{Classical 2D MUSIC Algorithm}

The MUSIC algorithm is widely used for AoA estimation through eigenvalue decomposition. Based on the model in~\eqref{YT}, the covariance matrix of the received signals is given by
\begin{equation} \label{RY}
    \begin{aligned}
        \boldsymbol{R}&=E [ \boldsymbol{Y} \boldsymbol{Y}^\mathrm{H} ] \\  
                                                      &= \boldsymbol{A} \boldsymbol{R}_{ss} \boldsymbol{A}^\mathrm{H} + \sigma^2\boldsymbol{I},
   \end{aligned}
\end{equation}
where $\boldsymbol{R}_{ss} = E[\boldsymbol{S} \boldsymbol{S}^\mathrm{H}]$ is the correlation matrix of the source signals. The eigenvectors of $\boldsymbol{R}$ associated with the largest $D$ eigenvalues span the signal subspace $\boldsymbol{E}_S$, while the remaining eigenvectors span the noise subspace $\boldsymbol{E}_N$. The 2D MUSIC AoA pseudo-spectrum is defined as
\begin{equation} \label{PR}
    \begin{aligned}
        \boldsymbol{P_{M}}(\theta,\phi)&= \frac{\boldsymbol{a}^H(\theta,\phi) \boldsymbol{a}(\theta,\phi)}{\boldsymbol{a}^H(\theta,\phi) \boldsymbol{E}_N^{} \boldsymbol{E}^H_N \boldsymbol{a}(\theta,\phi)}.
   \end{aligned}
\end{equation}
The angles corresponding to the peaks in this pseudo-spectrum provide estimates of the directions of the incident signals.

\subsection{I-SSMUSIC for 3D AoA}\label{III-1}

In contrast to 2D AoA estimation, 3D AoA estimation demands significantly higher computational complexity. Moreover, 3D localization tasks are further challenged by the increased severity of multipath propagation. A well-known limitation of subspace-based methods is their degraded performance in the presence of correlated sources, primarily due to rank deficiency in the covariance matrix. A notable solution to mitigate this issue is the spatial smoothing technique.

We now present an improved MUSIC algorithm with 2D spatial smoothing, referred to as I-SSMUSIC, designed for URAs. Based on~\eqref{YT}, the $(m_1,m_2)$-th smoothed subarrays of size $M_{1} \times M_{2}$ is formally expressed as
\begin{equation} \label{YS} 
    \boldsymbol{Y}_{m_1 m_2} = \boldsymbol{A}_1 \boldsymbol{D}_x^{m_1-1} \boldsymbol{D}_y^{m_2-1} \cdot \boldsymbol{S} + \boldsymbol{N}_{m_1 m_2},
\end{equation}
where 
\begin{equation} 
    \begin{aligned}
        \boldsymbol{D}_x & = \text{diag}{[u(\theta_1, \phi_1),\cdots,u(\theta_L, \phi_L)]}, \\
        \boldsymbol{D}_y & = \text{diag}{[v(\theta_1, \phi_1),\cdots,v(\theta_L, \phi_L)]}.
    \end{aligned}
\end{equation}
Here $\boldsymbol{N}_{m_1 m_2}$ is the noise matrix at the $(m_1,m_2)$-th subarray and $\boldsymbol{A}_1 = \left[ \boldsymbol{a}_1(\theta_1,\phi_1) \ \boldsymbol{a}_1(\theta_2,\phi_2) \ \cdots \ \boldsymbol{a}_1(\theta_L,\phi_L) \right]$ is the steering matrix, where each $\boldsymbol{a}_1(\theta_l,\phi_l)$ is given by
\begin{equation} 
    \begin{aligned}
        & \boldsymbol{a}_1(\theta_l,\phi_l) = \boldsymbol{a}_{y,M_1}\left(\theta_{l}, \phi_{l}\right) \otimes \boldsymbol{a}_{x,M_1}\left(\theta_{l}, \phi_{l}\right), \\
        & \boldsymbol{a}_{x,M_1}(\theta, \phi) = \left[1 \quad u \quad \cdots \quad u^{M_1-1}\right]^{\top},\\
        & \boldsymbol{a}_{y,M_2}(\theta, \phi) = \left[1 \quad v \quad \cdots \quad v^{M_2-1}\right]^{\top}.   
    \end{aligned}
\end{equation}

\begin{figure}
    \centerline{\includegraphics[width=0.82\columnwidth]{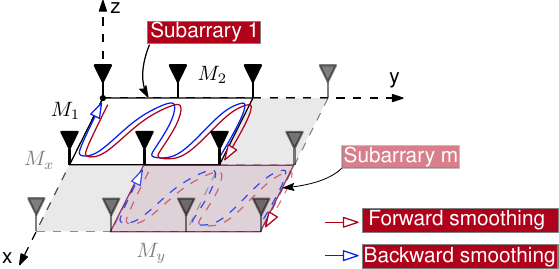}} 
    \caption{I-SSMUSIC of URA with forward-backward spatial smoothing applied to each subarray.}
    \label{URA-SS} 
\end{figure}

Using~\eqref{YS}, we can reformulate the expression in~\eqref{RY}. The covariance matrix of the $(m_1,m_2)$-th subarray is therefore given by
\begin{equation} 
    \begin{aligned}
        \boldsymbol{R}_{m_1 m_2}^f & = \boldsymbol{A}_1 \boldsymbol{D}_x^{m_1-1} \boldsymbol{D}_y^{m_2-1} \boldsymbol{R}_{ss} \left(\boldsymbol{D}_y^{m_2-1}\right)^\mathrm{H} \\
        &\quad \quad \quad \quad \quad \quad \quad \quad \times \left(\boldsymbol{D}_x^{m_1-1}\right)^\mathrm{H} \boldsymbol{A}_1^\mathrm{H} + \sigma^2 \boldsymbol{I}.
    \end{aligned}
\end{equation}
In the spatial smoothing scheme, the forward smoothed covariance matrix $\boldsymbol{R}^f$ is obtained by averaging the covariance matrices of all forward subarrays, yielding
\begin{equation}\label{E:R_F}
    \begin{aligned}
        \boldsymbol{R}^f = \frac{1}{H_x H_y} \sum^{H_x}_{m_1=1}\sum^{H_y}_{m_2=1} \boldsymbol{R}_{m_1 m_2}^f = \boldsymbol{A}_1 \boldsymbol{R}_s^f \boldsymbol{A}_1^\mathrm{H} + \sigma^2 \boldsymbol{I},
    \end{aligned}
\end{equation}
where $H_x = M_x - M_1 + 1$ and $H_y = M_y - M_2 + 1$. Similarly, we denote the forward-smoothed source covariance matrix by $\boldsymbol{R}_s^f$, which is defined by 
\begin{equation}
    \begin{aligned}
        \boldsymbol{R}_s^f &= \frac{1}{H_x H_y} \sum^{H_x}_{m_1=1}\sum^{H_y}_{m_2=1} \boldsymbol{D}_x^{m_1-1} \boldsymbol{D}_y^{m_2-1} \boldsymbol{R}_{ss}  \\
        & \quad \quad \quad \quad \quad \quad \quad \quad \quad \times \left(\boldsymbol{D}_y^{m_2-1}\right)^\mathrm{H} \left(\boldsymbol{D}_x^{m_1-1}\right)^\mathrm{H}.
    \end{aligned} 
\end{equation} 
The spatially smoothed covariance matrix enables the application of eigenstructure-based methods for AoA estimation, even in the presence of coherent signals.

One limitation of the spatial smoothing algorithm is its tendency to reduce the effective array aperture, which may degrade sensing performance~\cite{kotaru2015spotfi}. To mitigate this issue, we introduce a forward-backward spatial smoothing scheme for URAs, as illustrated in Fig.~\ref{URA-SS}. This bidirectional smoothing approach preserves the aperture size by exploiting the conjugate symmetry property of the covariance matrix.

Mathematically, the forward-backward spatially smoothed covariance matrix is expressed as
\begin{equation} \label{Rx}
    \begin{aligned}
        \boldsymbol{R}_{\boldsymbol{X}} = \frac{1}{2} \left (\boldsymbol{R}^f + \boldsymbol{I}_{\boldsymbol{v}} {\left (\boldsymbol{R}^f\right )}^{*} \boldsymbol{I}_{\boldsymbol{v}}\right ),
    \end{aligned}
\end{equation} 
where $\left (\boldsymbol{R}^f\right )^{*}$ is the conjugate for matrix $\boldsymbol{R}^f$, and 
\begin{equation} \label{Iv} 
    \begin{aligned}
        \boldsymbol{I}_{\boldsymbol{v}} = \begin{bmatrix} 
            0 & \cdots & 0 & 1 \\
            0 & \cdots & 1 & 0 \\
            \vdots & \reflectbox{$\ddots$} & \vdots & \vdots \\ 
            1 & \cdots & 0 & 0 \\ 
        \end{bmatrix}_{M \times M}.
    \end{aligned} 
\end{equation} 
By computing the pseudo-spectrum in \eqref{PR} using this smoothed covariance matrix, we enable accurate estimation of correlated signals while mitigating the effects of rank deficiency.

By examining \eqref{E:R_F} and \eqref{Rx}, we observe that the number of forward-only smoothed subarrays, denoted by $H$, determines the maximum number of resolvable correlated sources, whereas forward-backward smoothing effectively doubles this limit to $2H$. In typical indoor environments, where the number of multipath components is usually fewer than five~\cite{gjengset2014phaser,kotaru2015spotfi}, a single forward-backward smoothing operation ($H=2$) can decorrelate signals from up to four distinct angles.

\begin{figure}
    \centerline{\includegraphics[width=1\columnwidth]{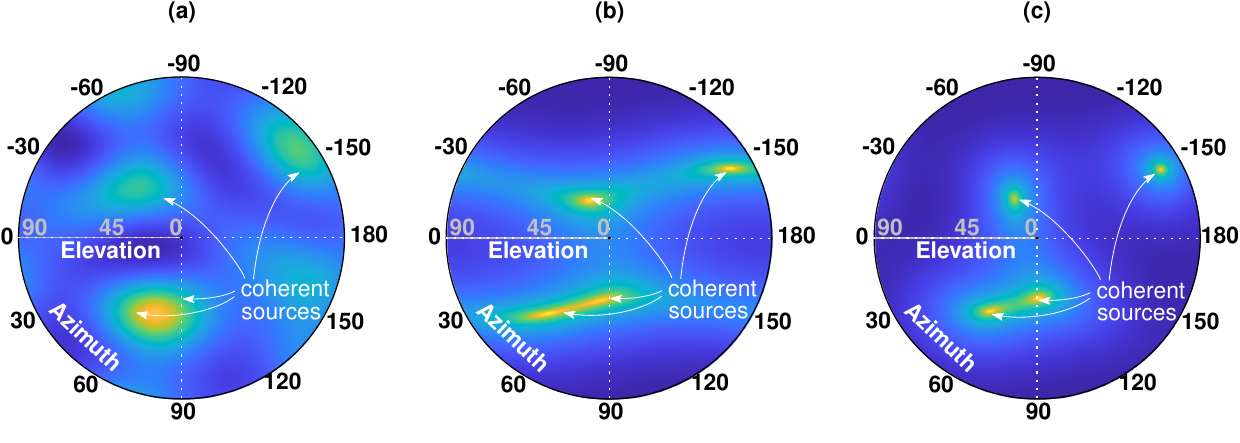}} 
    \caption{Spatial spectrum of four correlated sources generated using a $3 \times 4$ antenna array. (a), (b) and (c) show the 2D spatial spectrums computed by MUSIC, SS-MUSIC and I-SSMUSIC, respectively. } 
    \label{IMUSIC_MUSIC_SSMUSIC}  
\end{figure}

We now present a comparative evaluation of conventional MUSIC, MUSIC with forward-only spatial smoothing (SS-MUSIC), and the proposed I-SSMUSIC for estimating the angles of four correlated signals under identical conditions. The URA consists of $3 \times 4$ antennas. Four correlated signal sources emit continuous signals with an SNR of 15 dB, arriving from the following angles: $(21.8 ^\circ, 90 ^\circ)$, $ (32^\circ, 56^\circ)$, $ (15^\circ, -60^\circ)$ and $ (60^\circ, -150^\circ)$, respectively. The spatial spectra are illustrated in Fig.~\ref{IMUSIC_MUSIC_SSMUSIC}, from which it is evident that the proposed I-SSMUSIC outperforms the other methods. The estimated AoAs using I-SSMUSIC are $(21.8 ^\circ, 90.8 ^\circ)$, $ (32.4^\circ, 57.2^\circ)$, $ (16.4^\circ, -59.6^\circ)$ and $ (60.2^\circ, -150.6^\circ)$, respectively. In comparison, while SS-MUSIC is capable of estimating correlated signals, it exhibits notably lower resolution. Its estimated AoAs are $(22.8 ^\circ, 82.2 ^\circ)$, $ (37.2^\circ, 50.8^\circ)$, $ (15.2^\circ, -62^\circ)$ and $ (58.8^\circ, -149.6^\circ)$. The standard MUSIC algorithm, by contrast, fails to resolve the correlated sources, resulting in an ambiguous and inaccurate AoA spectrum.

\begin{figure*}
    \centerline{\includegraphics[width=2\columnwidth]{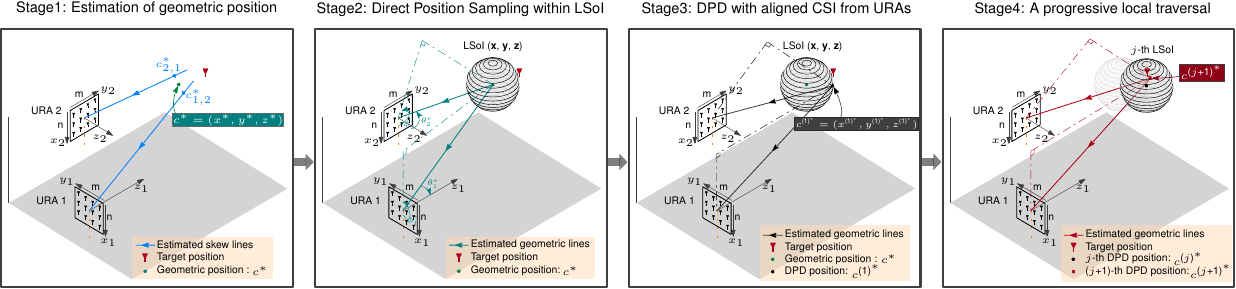}} 
    \caption{Collaborative 3D DPD localization. The initial independent angle estimations from two URAs yield two skew lines. A geometric-based 3D positioning method is employed to determine a geometric position (GP). The geometric center serves as the anchor point for defining a localized space of interest (LSoI). A joint 3D positioning algorithm is applied to further refine the source location.} 
    \label{2URA_skewline} 
\end{figure*}

\subsection{Closest Geometric Point Estimation} \label{Multiple3DAoA} 

With AoA estimations obtained from multiple URAs distributed across space, the specific location of the signal source can be determined. Ideally, the estimated AoA vectors intersect at the true position of the source. However, due to measurement errors, a robust closest-point estimation algorithm is required to approximate the actual point of intersection. The proposed geometric positioning (GP) method first identifies the closest points between each pair of AoAs, as illustrated in Stage 1 of Fig.~\ref{2URA_skewline}. The final position estimate is then computed as the mean of these closest points.

Let $l_i$ denote the estimated arrival ray associated with the $i$-th URA. Each ray can be represented by a parametric equation of the form
\begin{equation} \label{geometric1}
    \begin{aligned}
    \left\{\begin{matrix}
        \boldsymbol{r}_1 = \boldsymbol{c}_1 + t_1 \boldsymbol{d}_1, \\ 
        \vdots \\
        \boldsymbol{r}_i = \boldsymbol{c}_i + t_i \boldsymbol{d}_i, \\ 
        \vdots \\
        \boldsymbol{r}_u = \boldsymbol{c}_u + t_u \boldsymbol{d}_u, \\ 
    \end{matrix}\right.    
    \end{aligned}
\end{equation} 
where $\boldsymbol{c}_i \in \mathbb{R}^3$ denotes the center of the $i$-th URA, and $\boldsymbol{d}_i \in \mathbb{R}^3$ is the direction vector of the arrival ray $l_i$. To identify the vector $\mathbf{t}_{h,i} = [t_h \ t_i]^\top$ that best approximates the intersection of the $h$-th and $i$-th AoA rays, we solve the following equation
\begin{equation} \label{t1t2}
    \begin{bmatrix}
       -\mathbf{d}_h^\top \mathbf{d}_h & \mathbf{d}_i^\top \mathbf{d}_h \\
       -\mathbf{d}_h^\top \mathbf{d}_i & \mathbf{d}_i^\top \mathbf{d}_i
    \end{bmatrix}
    \begin{bmatrix}
    t_h \\
    t_i
    \end{bmatrix} =
    \begin{bmatrix}
    (\mathbf{c}_h - \mathbf{c}_i) \cdot \mathbf{d}_i \\
    (\mathbf{c}_h - \mathbf{c}_i) \cdot \mathbf{d}_h
    \end{bmatrix}.
\end{equation} 
If there is no exact intersection, the least-squares solution $\mathbf{t}^* = [t_{h,i}^* \ t_{i,h}^*]^\top$ determines the pair of closest points on the two rays. The coordinates of these points are given by
\begin{equation} \label{point1}
    \begin{aligned}
    \mathbf{c}_{h,i}^* = \boldsymbol{c}_h + t_{h,i}^* \boldsymbol{d}_h,
    \mathbf{c}_{i,h}^* = \boldsymbol{c}_i + t_{i,h}^* \boldsymbol{d}_i. 
    \end{aligned}
\end{equation} 
Finally, the estimated position of the source based on all $u$ URAs is computed as
\begin{equation} \label{pointstar}
    \mathbf{c}^* = \frac{1}{u(u-1)}\sum_{h=1}^{u-1} \sum_{i=h+1}^{u} (\mathbf{c}_{h,i}^* + \mathbf{c}_{i,h}^*) = [x^*, y^*, z^*]. 
\end{equation} 

\section{Collaborative 3D Direct Position Determination} \label{Joint 3D Direct Position Determination} % wavelength level 

For the previously described closest geometric point approach, collaboration is performed at the level of estimated AoAs, as the involved URAs are not synchronized with each other. Given that signal synchronization among elements within each array has now been implemented, a natural question arises: can this synchronization mechanism be further extended to the inter-array level to enable greater cooperative gains? In this section, we develop an inter-array synchronization framework designed to facilitate direct position determination (DPD)~\cite{tirer2015high,wang2022computationally}. Unlike the preceding closest point estimation method, DPD bypasses intermediate parameter estimation, such as AoA, and instead computes the source position directly in a single step.

To reduce the spatial sampling overhead of the proposed DPD algorithm, we first employ the I-SSMUSIC and closest point estimation approaches to define a compact localized space of interest (LSoI). By discretizing the LSoI, we derive a measurement model that characterizes the observation process across multiple synchronized URAs. Synchronization among these arrays is achieved by measuring phase differences relative to a common reference signal. Once synchronization is established, the distributed URAs effectively form a virtual large-scale array, enabling the computation of the MUSIC pseudo-spectrum at the spatial sampling points within the LSoI. To further expand the LSoI and enhance estimation fidelity, we introduce a progressive local traversal strategy. The overall process is illustrated in Fig.~\ref{2URA_skewline}.

For simplicity, the LSoI is configured as a sphere of radius $R$, centered at the closest geometric point $\boldsymbol{c}^{*}$ estimated via the I-SSMUSIC algorithm. The sphere is discretized with a voxel size of $q$, yielding a total of $K$ spatial sampling points. To construct the steering matrix for the virtual large-scale array, it is necessary to convert the Cartesian coordinates of each sampling point into the corresponding azimuth and elevation angles relative to each URA. Accordingly, the steering matrix for the $i$-th URA is expressed as 
\begin{equation}
\boldsymbol{A}_{\text{URA}_i}(\boldsymbol{x},\boldsymbol{y},\boldsymbol{z})=\left[ \boldsymbol{a}_i(x_1,y_1,z_1) \ \cdots \ \boldsymbol{a}_i(x_K,y_K,z_K) \right],
\end{equation} 
where $\boldsymbol{a}_i(x_k,y_k,z_k)$ denotes the steering vector associated with the sampling point $(x_k,y_k,z_k)$ with respect to the $i$-th URA.

For synchronization among multiple URAs, an inter-array phase alignment procedure is applied to measure the phase difference $\bigtriangleup \hat{\gamma}_i\in \mathbb{C}$ between the reference signals of $\text{URA}_1$ and $\text{URA}_i$. The received signal matrix for $\text{URA}_i$ is then adjusted as $\hat{\boldsymbol{Y}}_{\text{URA}_i} = \bigtriangleup \hat{\gamma}_i \cdot \boldsymbol{Y}_{\text{URA}_i}$. Consequently, the joint received signal model across two URAs can be expressed as
\begin{equation} \label{URA12}
    \begin{bmatrix} 
        \hat{\boldsymbol{Y}}_{\text{URA}_1} \\
        \hat{\boldsymbol{Y}}_{\text{URA}_2}
    \end{bmatrix} 
    =
    \begin{bmatrix}
        \boldsymbol{A}_{\text{URA}_1} \\
        \boldsymbol{A}_{\text{URA}_2} \cdot \bigtriangleup\boldsymbol{\Gamma}
    \end{bmatrix}
    \cdot \boldsymbol{S} + \boldsymbol{N},
\end{equation} 
where $\bigtriangleup \boldsymbol{\Gamma} = \text{diag}\{\bigtriangleup \gamma_{1}, \cdots, \bigtriangleup \gamma_{K} \}$  is a diagonal matrix encoding the inter-array phase offsets. 

The key advantage of this approach is that the signal subspace of the synthesized virtual array is spanned by the dominant eigenvectors of the aggregated covariance matrix. The subspace-based MUSIC algorithm is then applied to estimate the 3D LSoI fused spectrum, as illustrated in Stage 3 of Fig.~\ref{2URA_skewline}.

\begin{algorithm}
    \caption{Collaborative 3D DPD Algorithm (DPD2URA) }
    \label{algorithm_fusion} 
    \KwIn{Phase calibrated CSI $\{y_i\}|^2_{i=1}$, number of antennas $M$, number of AoA grid points $L$, steering matrix $\boldsymbol{A}$, the number of grid point $K$.}  
    \KwOut{ Location of the source $[\hat{x}, \hat{y}, \hat{z}]$.}  
    \For{ \rm{each URA} $i \in 1,2$} 
    {
        % Calibrate the CSI phase within $i$-th URA\;
        Calculate the covariance matrix $\boldsymbol{R}$ and $\boldsymbol{R}_{\boldsymbol{X}}$ using~\eqref{Rx} \;
        Construct matrix $\boldsymbol{E}_{i,N}$ using I-SSMUSIC\;
        \For{$l = 1, \ldots, L$}
        {
             $P_{I-M}\left( \theta_l, \phi_l \right) = \frac{\boldsymbol{a}_i(\theta_l, \phi_l)\boldsymbol{a}_i(\theta_l, \phi_l)}{\boldsymbol{a}_i(\theta_l,\phi_l) \boldsymbol{E}_{i,N} \boldsymbol{E}_{i,N} \boldsymbol{a}_i(\theta_l, \phi_l)}$\;
        }
        Find the peaks of AoA spectrum\; 
    }
    Obtain skew lines and compute the geometric position $\boldsymbol{c}^*$ using~\eqref{geometric1} to~\eqref{pointstar}\;
    Calculate the coordinate of grid points $\boldsymbol{c}_k^{(0)} = [x_{k},y_{k},z_{k}], k = 1, \cdots, K$\;
    Set steering matrix $\boldsymbol{A}_{\text{URA}_1}$ and $\boldsymbol{A}_{\text{URA}_2}$ and obtain the $\hat{\boldsymbol{Y}}_{\text{URA}_1}$ and $\hat{\boldsymbol{Y}}_{\text{URA}_2}$ by aligning the phase of $\text{URA}_1$ and $\text{URA}_2$ using~\eqref{URA12}\; 
    Construct matrix $\boldsymbol{E}_N^{syn}$ using MUSIC\; 
    \Repeat{\rm{convergence criterion is met}}
    {
    \For{$k = 1, \ldots, K$} 
    {
         $P_{M}\left(x_{k}^{(j+1)},y_{k}^{(j+1)},z_{k}^{(j+1)}\right) = 
         \frac{\boldsymbol{a}\left(x_{k}^{(j)},y_{k}^{(j)},z_{k}^{(j)}\right) \boldsymbol{a}\left(x_{k}^{(j)},y_{k}^{(j)},z_{k}^{(j)}\right)}{\boldsymbol{a}\left(x_{k}^{(j)},y_{k}^{(j)},z_{k}^{(j)}\right) \boldsymbol{E}_N^{syn} \boldsymbol{E}_N^{syn} \boldsymbol{a}\left(x_{k}^{(j)},y_{k}^{(j)},z_{k}^{(j)}\right)}$\;
    }
    Obtain position of the source $\boldsymbol{c}^{(j+1)^*} = [\hat{x}^{(j+1)^*},\hat{y}^{(j+1)^*},\hat{z}^{(j+1)^*}]$ within $(j+1)$-th LSoI\; 
    Replace $\boldsymbol{c}^{(j)^*}$ with $\boldsymbol{c}^{(j+1)^*}$\;
    Calculate the new coordinate of grid points $\boldsymbol{c}_{k}^{(j+1)}, k = 1, \cdots, K$\;
    Set $\boldsymbol{A}_{\text{URA}_1}^{(j+1)}$ and $\boldsymbol{A}_{\text{URA}_2}^{(j+1)}$\ using~\eqref{URA12}\;
    $j \leftarrow j+1$\; 
    }
    Select the peak with the largest value of the spectrum\;
    Obtain position of the source $[\hat{x},\hat{y},\hat{z}]$ through DPD.
\end{algorithm}

Finally, a progressive local traversal method is introduced to further expand the LSoI and enhance the fidelity of position estimation. Specifically, if an updated estimate $\boldsymbol{c}^{(j)^*} = [\hat{x}^{(j)^*},\hat{y}^{(j)^*},\hat{z}^{(j)^*}]$ is obtained, the current LSoI center is updated accordingly. The LSoI is then re-centered at $\boldsymbol{c}^{(j)^*}$, and the spatial sampling points are regenerated. The complete algorithmic procedure, referred to as DPD2URA, is outlined in Algorithm~\ref{algorithm_fusion}. In practical implementations, the search radius $R$ is set to 0.1 meters, and the voxel size $q$ is set to 0.005 meters. Empirical results demonstrate that the method typically converges within three iterations.

\section{Prototype Implementation and Experimental Environments}\label{Prototype}

\subsection{Hardware Settings}
Two computers, each equipped with an Intel 5300 NIC, are configured as the transmitter and receiver, respectively. The Linux CSI tool~\cite{Halperin_csitool} is employed to capture CSI for each packet operated in monitor mode. The Wi-Fi system operates at 5.2~GHz on channel 40 with a bandwidth of 40 MHz. The signal source transmits at a rate of 2000 packets per second using either an omnidirectional lollipop antenna or a directional horn antenna. To evaluate various localization scenarios, the source is mounted on a movable platform that enables precise control over its position and orientation.

We arrange 24 micro-antennas into two $3 \times 4$ URAs with an inter-element spacing of 0.54$\lambda$. These arrays function as receiving antennas for 3D AoA estimation. Each antenna is an omnidirectional lollipop antenna with a gain of 3 dBi. The inter-element spacing is determined based on an elevation angle of $\theta_L=60^\circ$ in the URA configuration. Each array collects 50 data packets before switching to the next. The complete data acquisition cycle takes approximately 0.1 seconds.

\subsection{Phase Offset Calibration}  
\begin{figure}%[htbp]
    \centerline{\includegraphics[width=0.8\columnwidth]{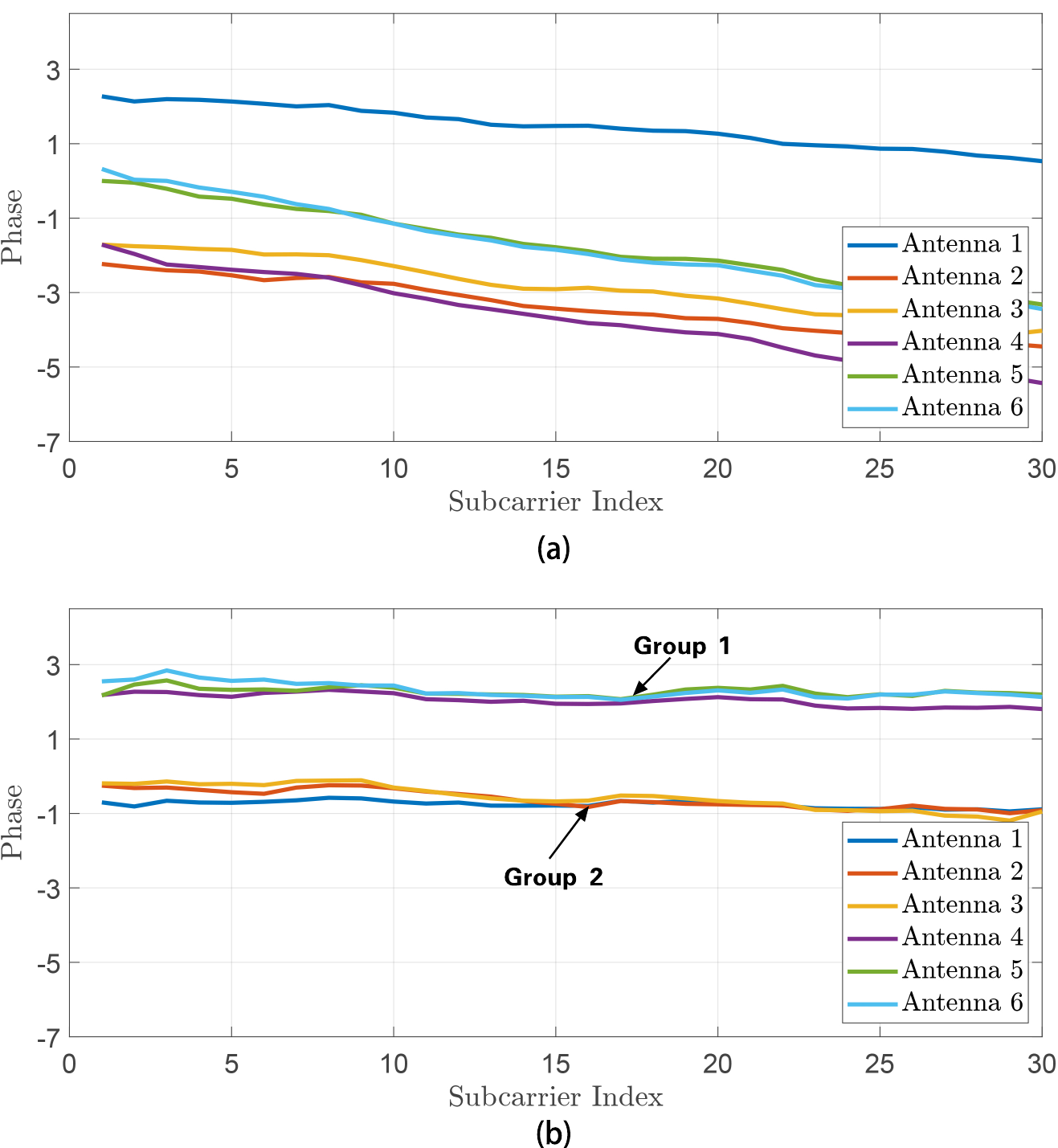}} 
    \caption{(a) Collected raw phase. (b) Calibrated phase.} 
    \label{phase_}  
\end{figure}

We present a visualization of two sets of phase data captured before and after phase calibration. Fig.~\ref{phase_}(a) shows the uncalibrated phase responses of Group 1 and Group 2, while Fig.~\ref{phase_}(b) illustrates the corresponding calibrated phases. As the signal source is positioned to the right of the URA on the same horizontal plane, phase differences arise primarily between groups, whereas intra-group phase variations remain negligible.

\subsection{Experimental Environments} 
\begin{figure}
    \centerline{\includegraphics[width=0.9\columnwidth]{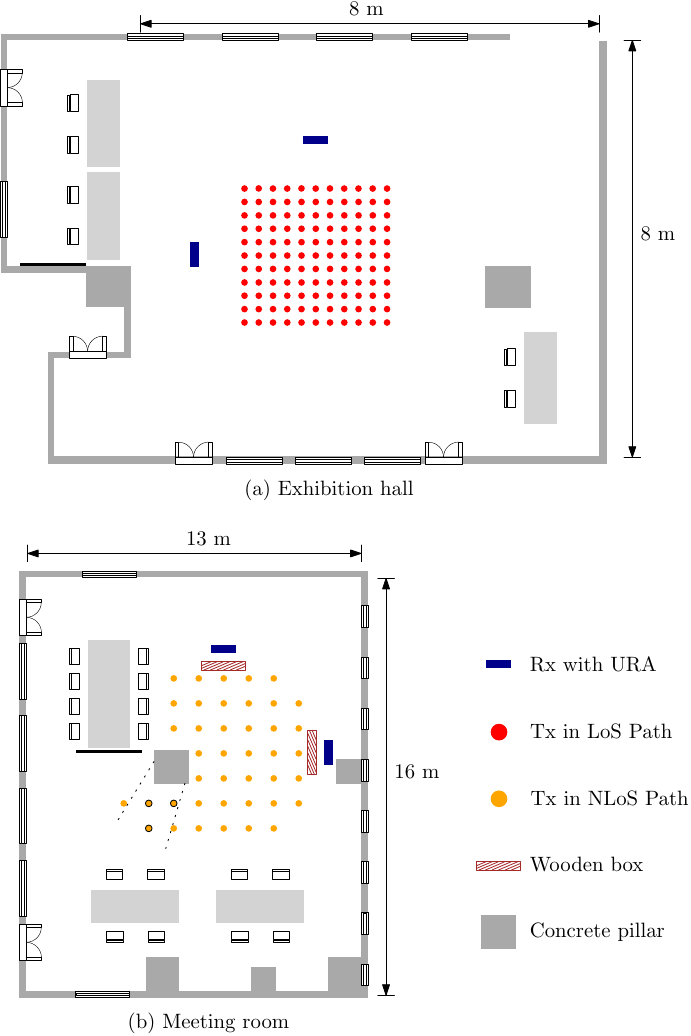}} 
    \caption{Layout of the experimental environments. } 
    \label{Testbed}  
\end{figure}

\begin{figure}
    \centerline{\includegraphics[width=1\columnwidth]{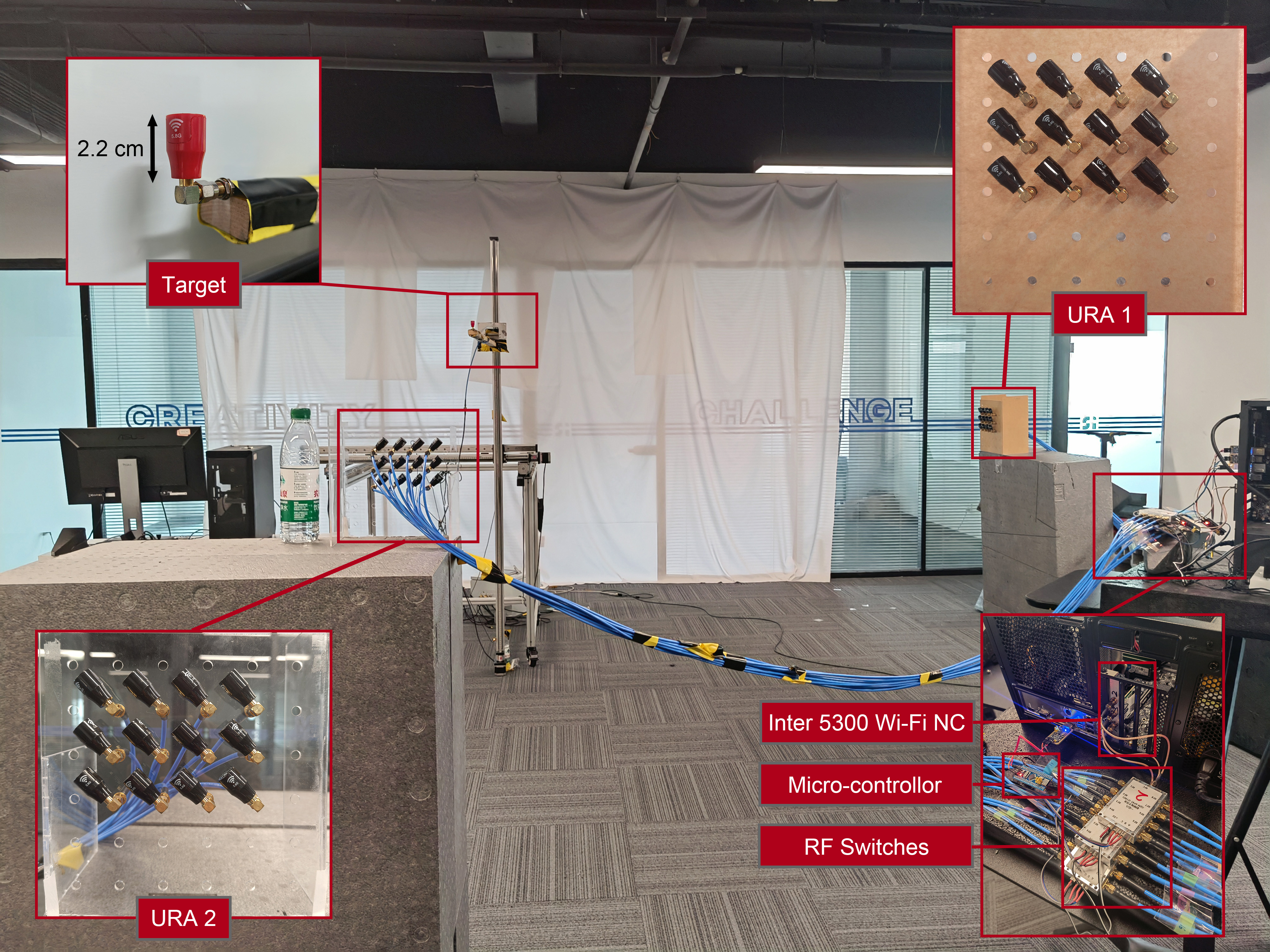}} 
    \caption{The implementation of WiCAL. The proposed position system consists of two URAs, connected via three SP8T switches to a single Intel 5300 Network Interface Card. Each URA is equipped with 12 micro-antennas.} 
    \label{URA_antenna_array}  
\end{figure}

We conduct experiments in two distinct indoor environments. The first is an exhibition hall measuring 8 m $\times$ 8 m $\times$ 3.5 m, which contains various indoor reflectors such as tables, chairs, concrete pillars, and metal obstacles, as shown in Fig.~\ref{Testbed}(a). This environment provides line-of-sight (LoS) conditions for signal propagation. A detailed view of the experimental setup is presented in Fig.~\ref{URA_antenna_array}.

The second environment is a larger and more complex indoor meeting room, measuring 13 m $\times$ 16 m, as depicted in Fig.~\ref{Testbed}(b). This setting presents significant challenges due to two primary factors: attenuation of line-of-sight (LoS) signals caused by wooden boxes manually placed in front of each URA, and multipath propagation induced by indoor furniture and structural concrete pillars.

\section{Performance Evaluation} \label{Evaluation}

\subsection{Evaluation of AoA Estimation Performance} 

\begin{figure}%[htbp!]
    \centerline{\includegraphics[width=0.65\columnwidth]{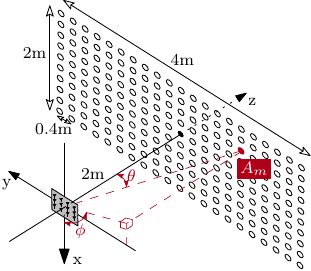}} 
    \caption{The planar region for 3D AoA estimation. A Tx device is mounted on a 3D moving platform, transmitting data at 0.4 m intervals over 210 angles.} 
    \label{SOI plane} 
\end{figure}

We begin by evaluating the performance of WiCAL in AoA estimation using a single URA. A planar testing region measuring 4 m $\times$ 2 m is positioned 2 m in front of the URA, allowing a maximum field of view of approximately 50 degrees. The signal source is placed at 210 distinct positions within this region, as illustrated in Fig.~\ref{SOI plane}.

\begin{figure} 
    \centerline{\includegraphics[width=1\columnwidth]{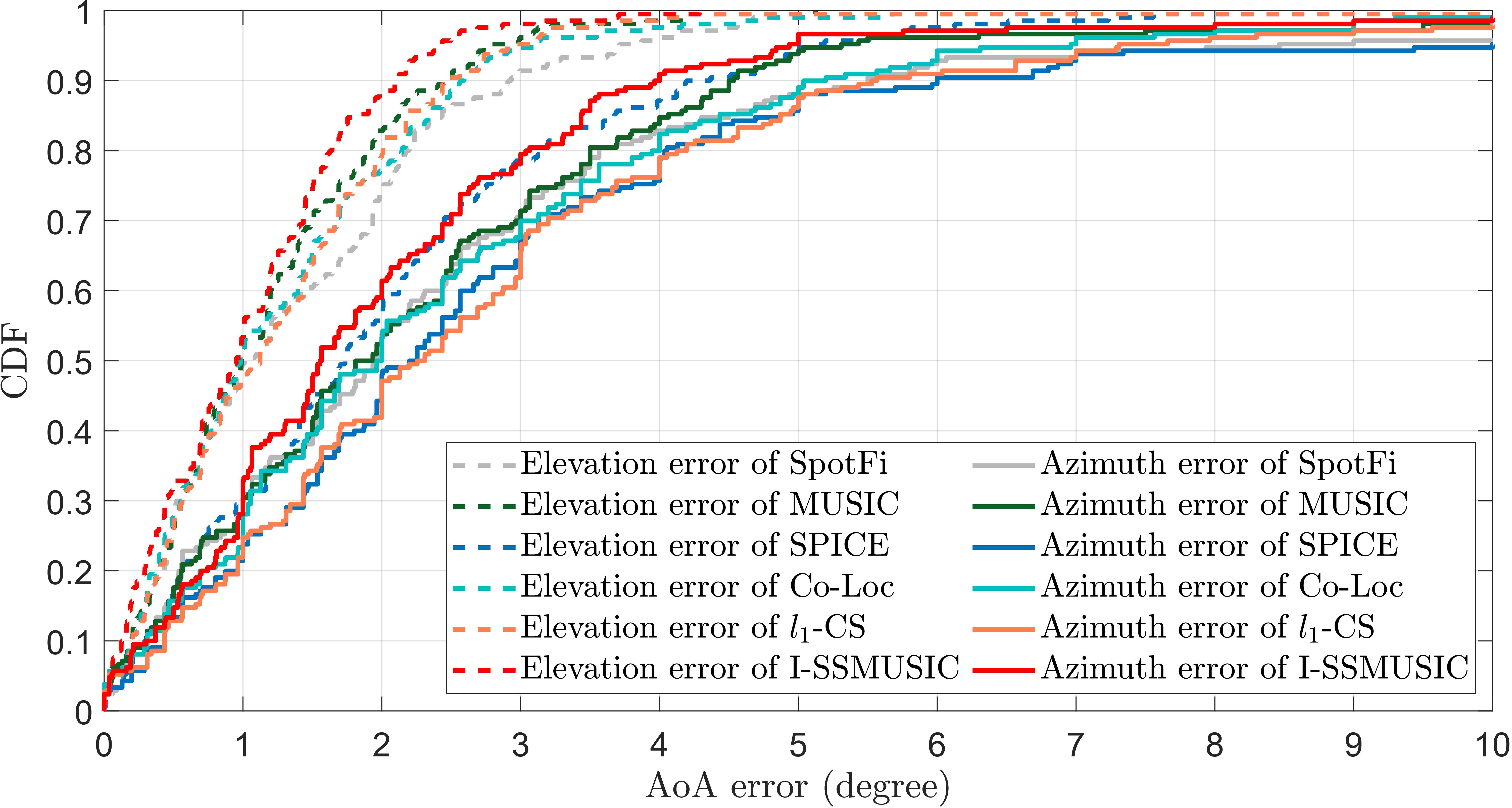}} 
    \caption{3D AOA estimation error.} 
    \label{AoA_error} 
\end{figure}

Five representative algorithms are implemented as baselines: MUSIC, SS-MUSIC~\cite{shan1985spatial}, SPICE~\cite{stoica2010spice}, $\ell_1$-norm compressed sensing (CS)~\cite{zhang20193d} and a non-parametric sparse recovery approach~\cite{yang2023multiple}. A comparison of AoA estimation error in both azimuth and elevation between the proposed I-SSMUSIC algorithm and these baseline methods is presented in Fig.~\ref{AoA_error}. The results show that the median errors in azimuth and elevation are $1^\circ$ and $1.5^\circ$, while the 90-th percentile errors reach $2.1^\circ$ and $4^\circ$. The I-SSMUSIC algorithm demonstrates significantly superior AoA estimation performance compared to the other algorithms.

\begin{figure}
  \centering
  \subfigure[]{
    \label{two_sources1}
    \includegraphics[width=.52\columnwidth]{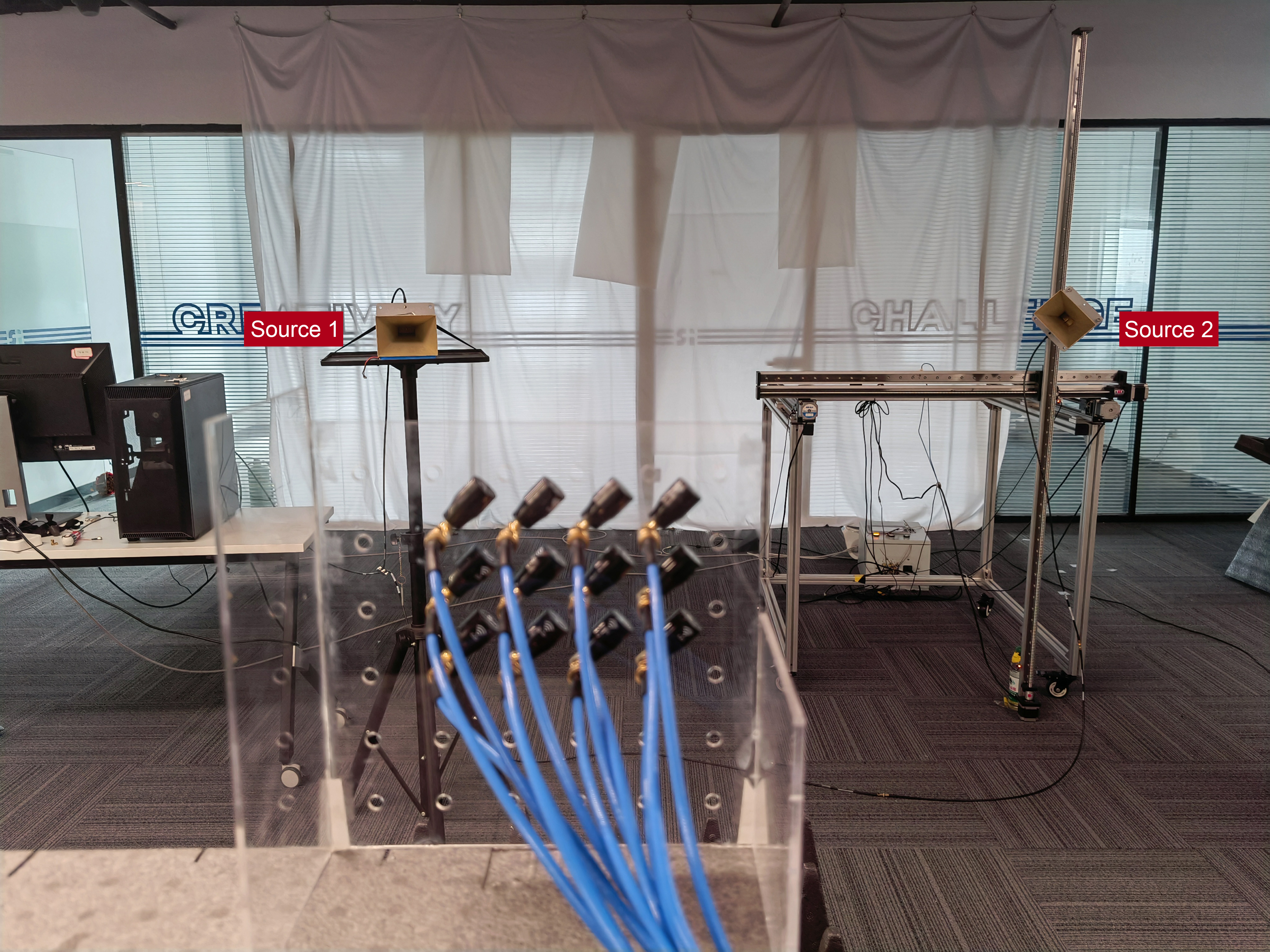}
  }
  \subfigure[]{
    \label{two_sources2}
    \includegraphics[width=.4\columnwidth]{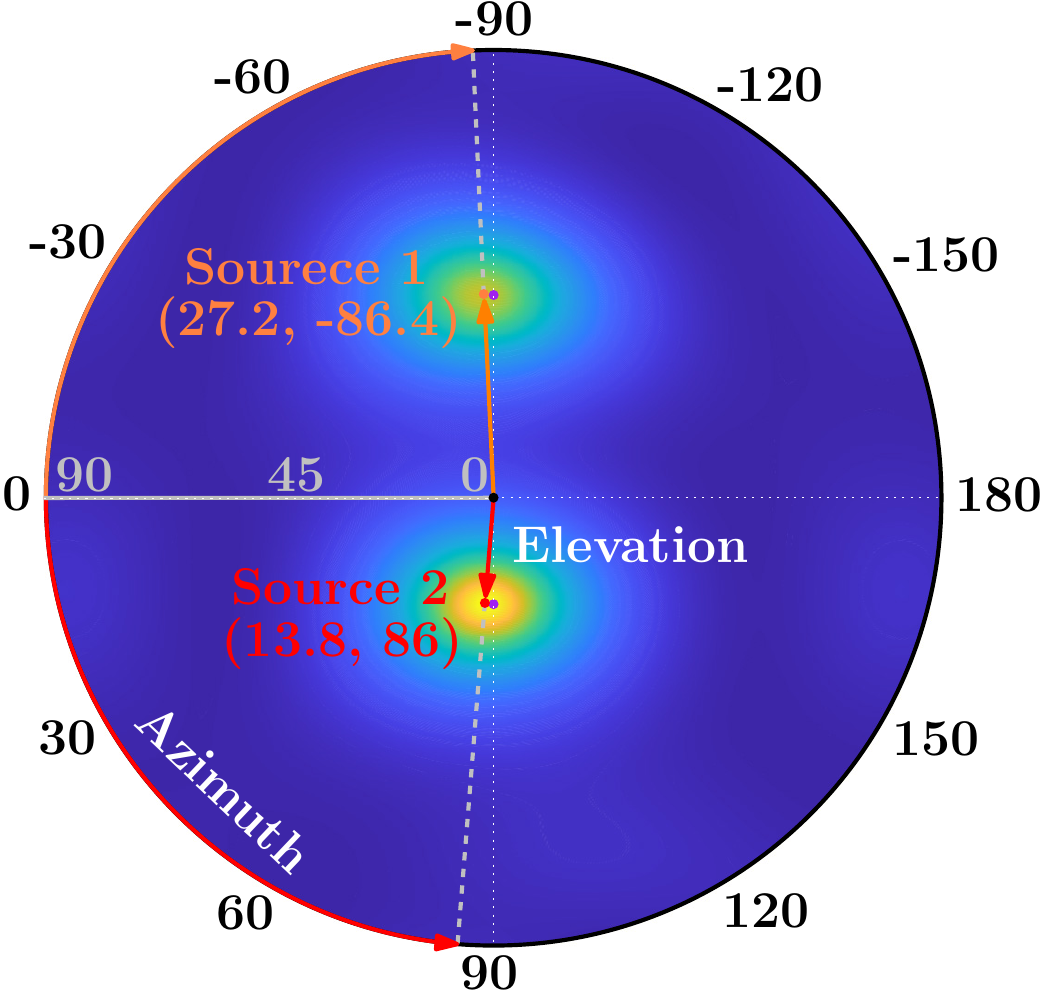}
  }
  \caption{(a) Experimental environment of two coherence sources. (b) The results obtained from I-SSMUSIC algorithm.}
  \label{F}
\end{figure}

We further investigate AoA estimation for multiple correlated sources. To generate two strong coherent signals, a power splitter connects the transmitter to two horn antennas via coaxial cables. Experimental environment is shown in Fig.~\ref{F}(a). The two horns are positioned at angular coordinates $(14^\circ, 90^\circ)$ and $(26.5^\circ, -90^\circ)$, respectively. The estimated results obtained from the I-SSMUSIC algorithm are visualized in Fig.~\ref{F}(b), where the estimated AoAs are $(13.8^\circ, 86^\circ)$ and $(27.2^\circ, -86.4^\circ)$, respectively. Both correlated signals are accurately localized, with estimation errors of  $0.45^\circ$ in elevation and $3.8^\circ$ in azimuth.

\subsection{3D Positioning via URAs} 

\begin{figure} 
    \centerline{\includegraphics[width=0.93\columnwidth]{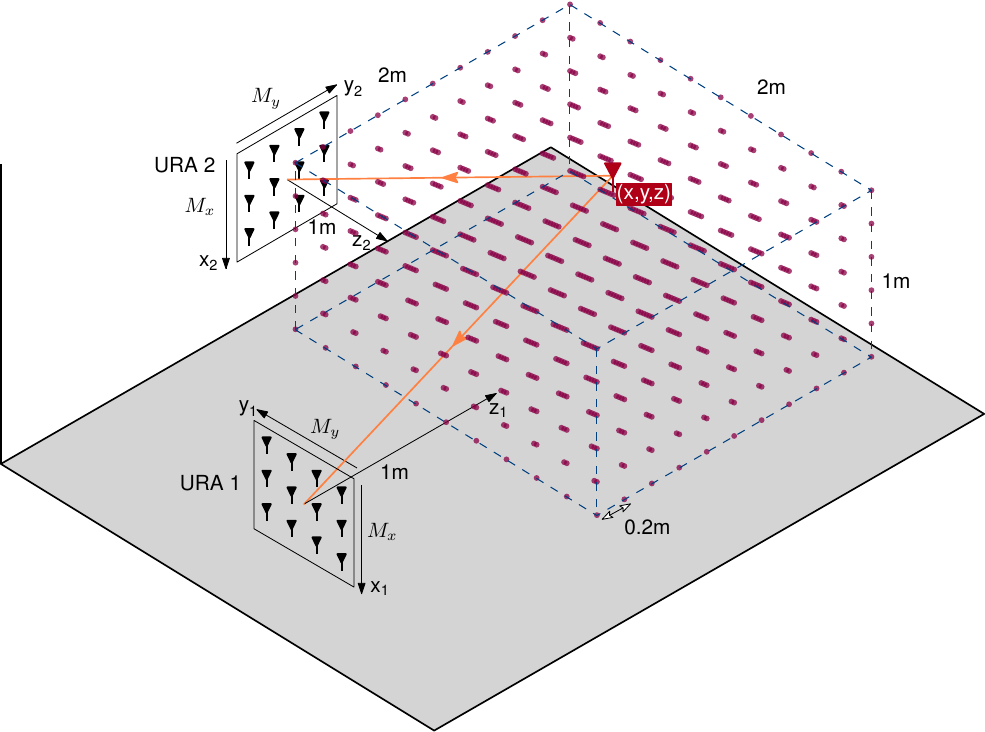}} 
    \caption{Schematic of the joint positioning of the two URAs working in tandem to localize a source in 3D space.} 
    \label{2URA-SOI} 
\end{figure}

\begin{figure}  
    \centerline{\includegraphics[width=1\columnwidth]{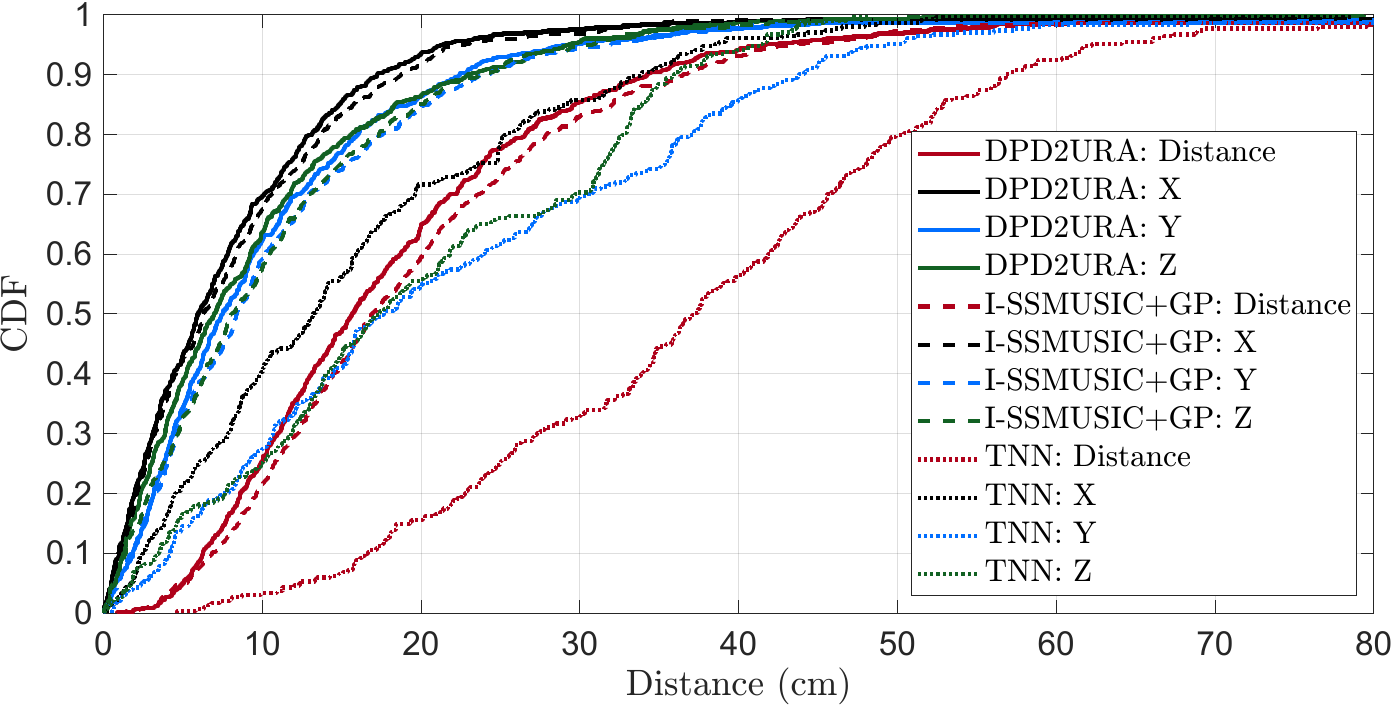}} 
    \caption{3D Localization results of the proposed DPD2URA, the geometric-based 3D positioning method (I-SSMUSIC+GP), and the AI-based TNN method.} 
    \label{distance error} 
\end{figure}

In this experiment, two $3 \times 4$ URAs are deployed to estimate the 3D position of the source. A total of $11 \times 11 \times 5$ sampling points are uniformly distributed within the region of interest, with a spacing of 0.2 m, as illustrated in Fig.~\ref{2URA-SOI}. The URAs are positioned at (0 m, 2 m, 1.17 m) and (2 m, 0 m, 1.17 m), respectively. We compare the performance of the proposed I-SSMUSIC+GP and DPD2URA methods with the AI-based multi-AoA spatial spectrum fusion localization network (TNN) proposed in iArk\cite{an2020general}. The plots in Fig.~\ref{distance error} reveal that the proposed DPD2URA algorithm achieves the lowest median localization error of 15.6 cm, with axis-wise errors of 6 cm along the $x$-axis, 7.5 cm along the $y$-axis, and 7 cm along the $z$-axis. When only the I-SSMUSIC+GP method is used, the accuracy decreases to a median error of 17 cm. The proposed joint DPD algorithm outperforms the geometric positioning method by 7.7\%. 

\begin{figure}%[!htbp]
    \centerline{\includegraphics[width=0.89\columnwidth]{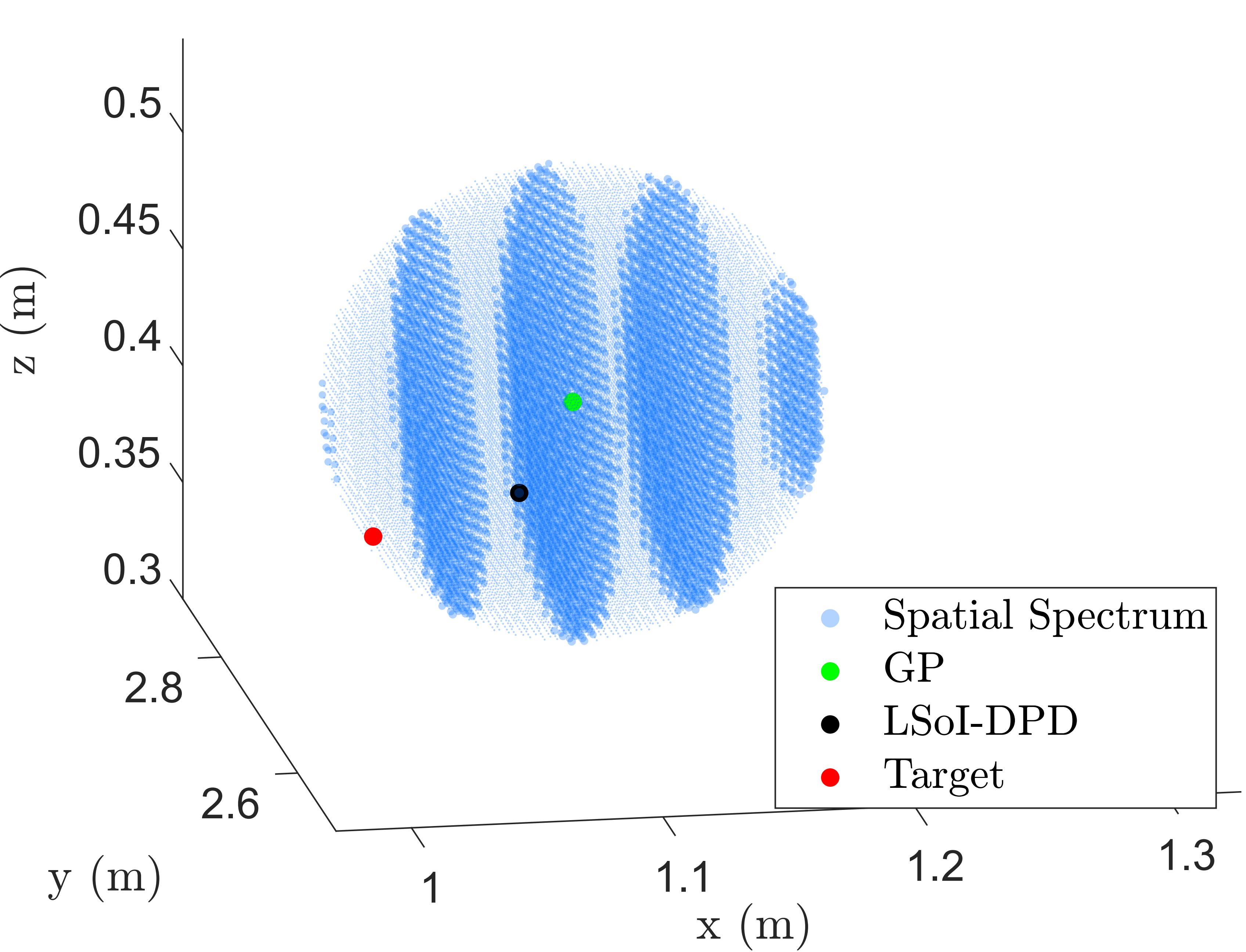}} 
    \caption{The spatial spectrum distribution obtained through the progressive local traversal search method provides refined localization results. } 
    \label{LSoI_error1} 
\end{figure}

Moreover, the estimations illustrated in Fig.~\ref{LSoI_error1} suggest that the progressive traversal search within the LSoI can further refine positioning accuracy. The blue points represent the spatial spectrum distribution obtained through joint estimation within the region, with larger dots corresponding to higher spectrum values. Following this progressive local traversal, the median positioning error is reduced to 15.6 cm, achieving a state-of-the-art result.~\footnote{The claim that this result surpasses those reported by iArk~\cite{an2020general} and SWAN~\cite{xie2018swan} is based on the numerical values presented in their original publications. However, this comparison may be limited in fairness, as their hardware platforms were not available for direct evaluation.} Notably, several dense blue regions appear in the spectrum, with each region spaced approximately at the wavelength scale. This observation suggests that synchronization influences the uniformity of the spectral distribution across the entire space, thereby confining spectral peaks to a limited number of regions.

\begin{figure} 
    \centerline{\includegraphics[width=1\columnwidth]{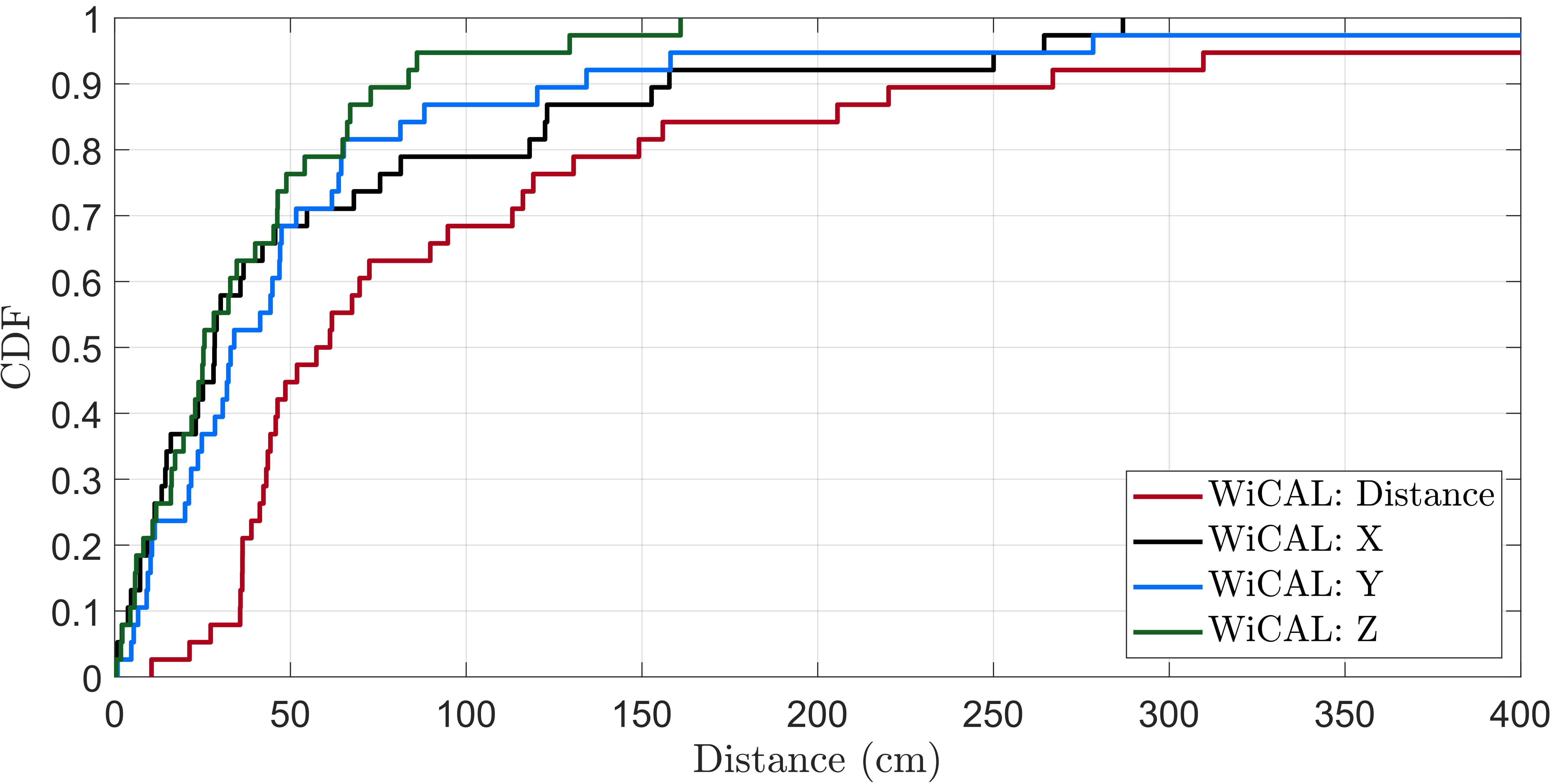}} 
    \caption{3D Localization errors in the NLoS settings.} 
    \label{dis_eor_meetingroom}  
\end{figure}

We further evaluate the performance of WiCAL in a larger and more challenging indoor meeting room, as shown in Fig.~\ref{Testbed}(b). Wooden boxes are manually placed in front of each URA to introduce LoS signal attenuation. This setup is designed to partially emulate non-line-of-sight (NLoS) conditions and to assess system performance under such circumstances. The results, presented in Fig.~\ref{dis_eor_meetingroom}, show that WiCAL achieves a median localization error of 0.57 m. These findings confirm that WiCAL maintains robust localization accuracy despite partial LoS blockage and rich multipath propagation.

Notably, under complete blockage conditions, where the transmitter is obstructed by concrete pillars, the localization accuracy deteriorates to approximately 3.2 m. This limitation is consistent with findings in existing Wi-Fi-based localization systems. Accurate positioning in NLoS environments with severe obstructions remains an open research challenge~\cite{li2023riscan}. A potential approach to mitigating this issue is to deploy additional URAs to circumvent NLoS blockages.

\subsection{Accuracy Evaluation of 3D Trajectory} 

\begin{figure} 
    \centerline{\includegraphics[width=0.86\columnwidth]{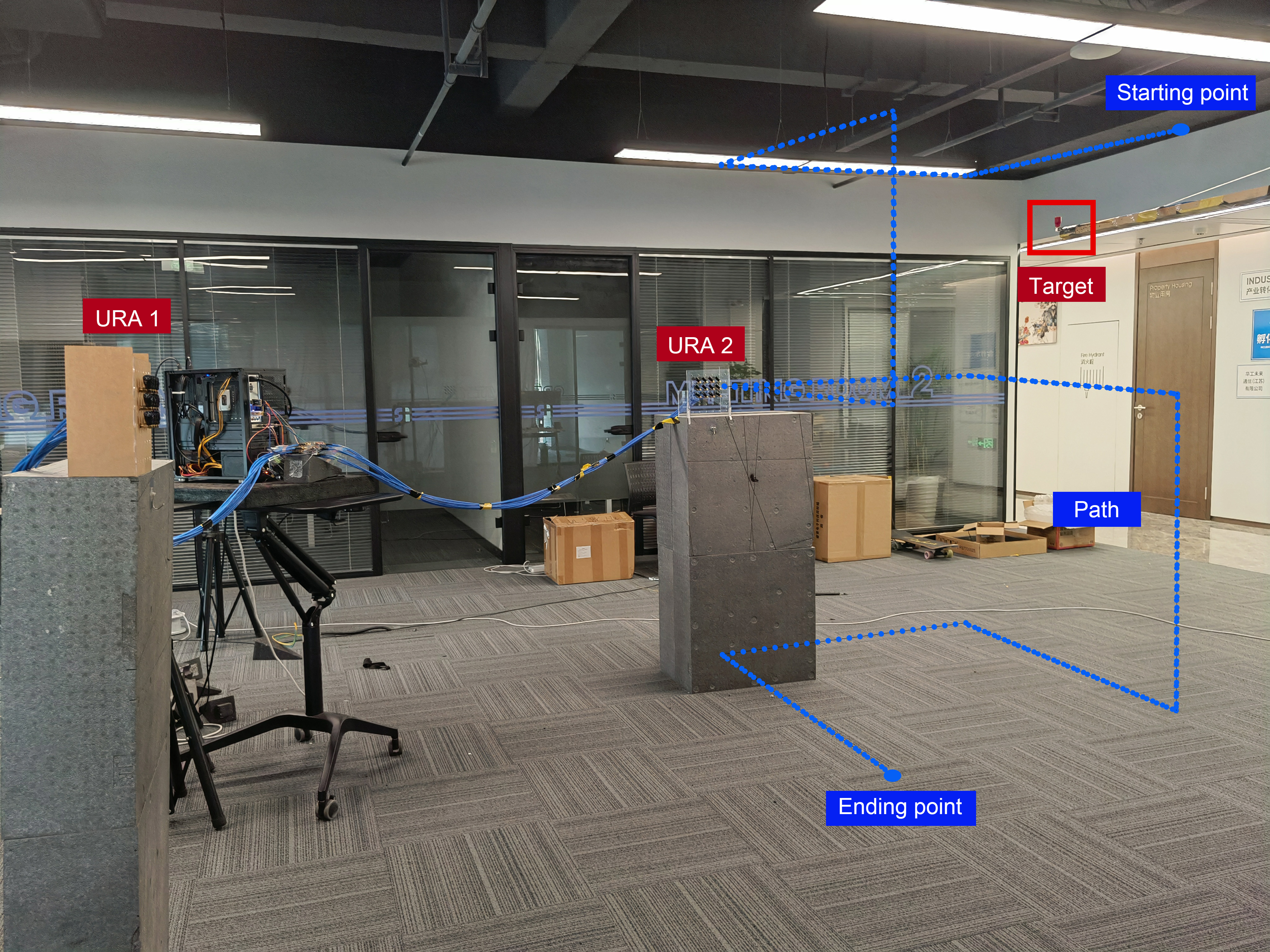}} 
    \caption{Experimental environment of 3D tracking.} 
    \label{2URA_tracking}  
\end{figure}

\begin{figure} 
    \centerline{\includegraphics[width=0.86\columnwidth]{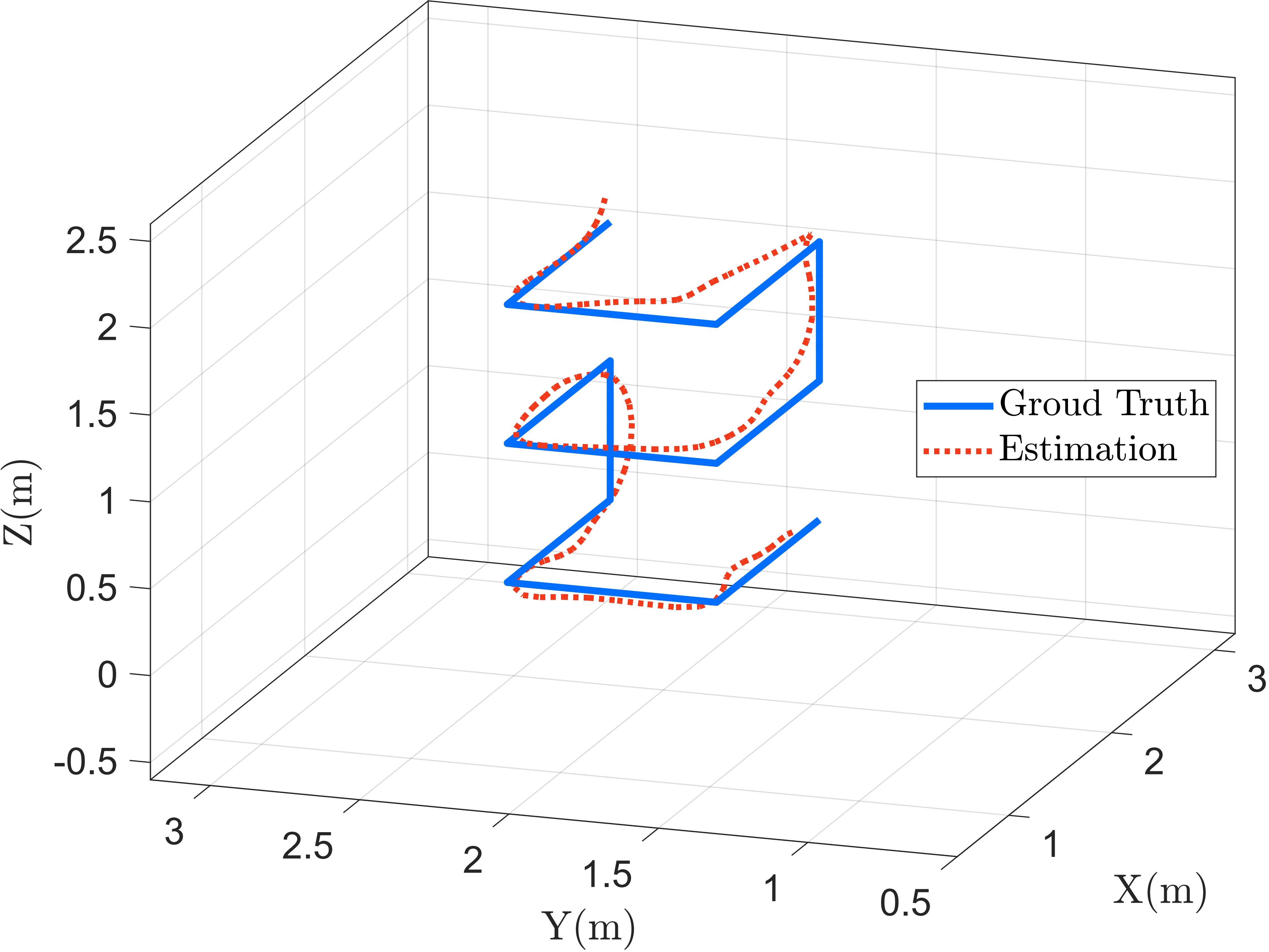}} 
    \caption{Tracking source with WiCAL. The red trajectory represent the estimated smoothed path by the WiCAL, while the blue line indicate the ground truth of the movement.} 
    \label{3D_tracking_Method_1}  
\end{figure}

To evaluate the performance of 3D trajectory reconstruction, an omnidirectional antenna mounted on a movable platform follows a predefined trajectory within a volume of approximately 1 m $\times$ 1 m $\times$ 2 m, as illustrated in Fig.~\ref{2URA_tracking}, where the path of the source is indicated by a blue line. Data from a total of 98 positions are recorded. Fig.~\ref{3D_tracking_Method_1} illustrates the reconstructed trajectory. A median-filter-based smoothing function is applied independently along the $x$-, $y$-, and $z$- axes to fit the discrete points into a continuous trajectory. The median error of the raw trajectory is 0.11 m, which is reduced to 0.075 m after smoothing.

\subsection{Impact of Array Size} 
\begin{figure}%[htbp!]
    \centerline{\includegraphics[width=0.7\columnwidth]{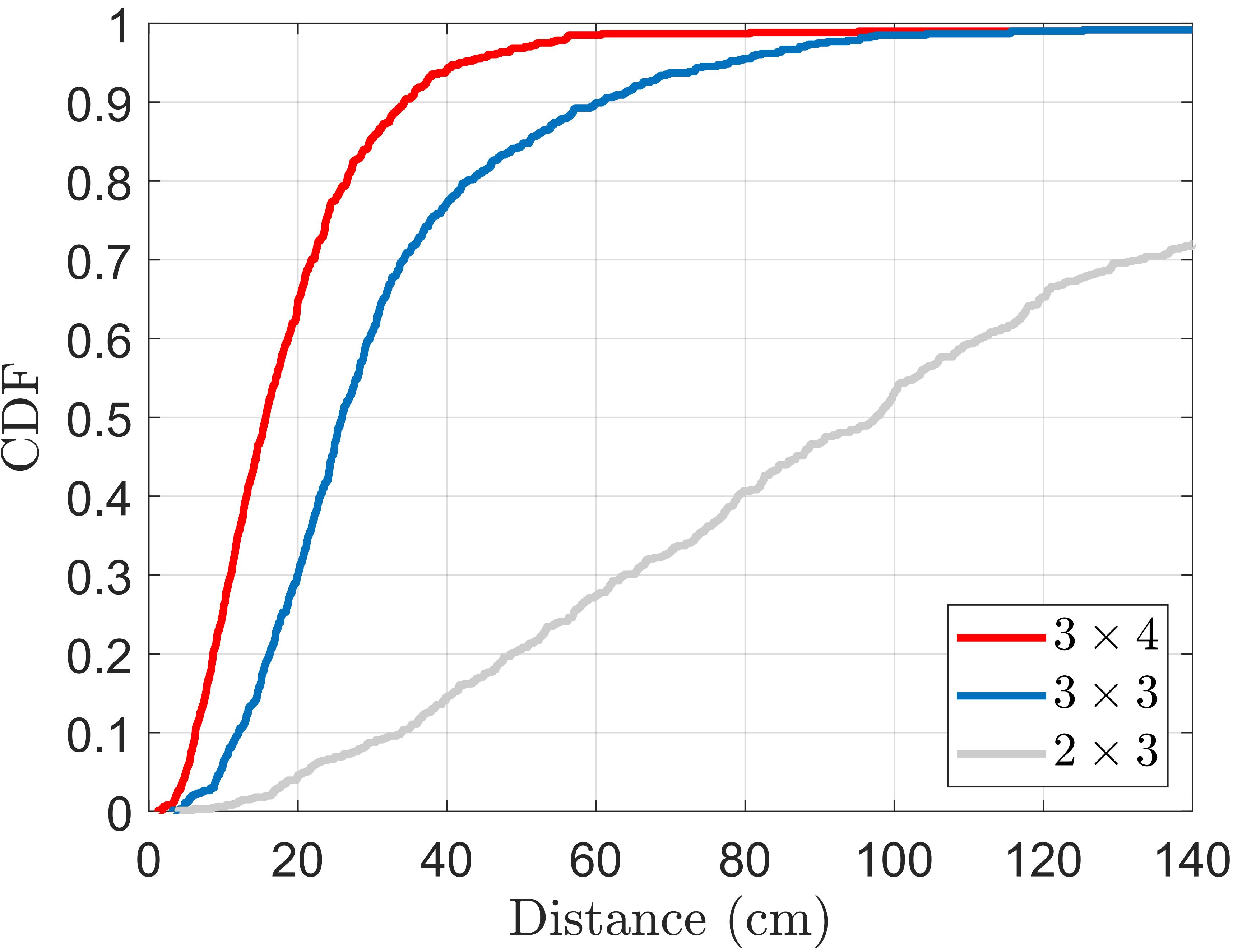}} 
    \caption{Impact of array size on 3D localization.} 
    \label{size}  
\end{figure}

We investigate the impact of array size on localization accuracy, by configuring the URA dimensions to $2 \times 3$, $3 \times 3$, and $3 \times 4$, respectively. The results, presented in Fig.~\ref{size}, indicate that the corresponding median localization errors are 97.5 cm, 26 cm, and 15.6 cm, respectively. These findings clearly demonstrate that increasing the antenna array size leads to a substantial improvement in positioning accuracy.

Indeed, by leveraging the proposed antenna extension and synchronization scheme, both the array size and the number of arrays can be further increased, thereby enhancing 3D localization accuracy. For instance, the current design utilizes a single tier of RF switches. Through cascaded connections, these switches can be extended to support a greater number of throws~\cite{xie2018swan}. This cascaded architecture enables seamless scalability, allowing WiCAL to efficiently accommodate larger antenna arrays and multiple distributed arrays.

\subsection{Time Complexity Comparison}\label{Time_Complexity_Comparison}
\begin{figure} 
    \centerline{\includegraphics[width=0.82\columnwidth]{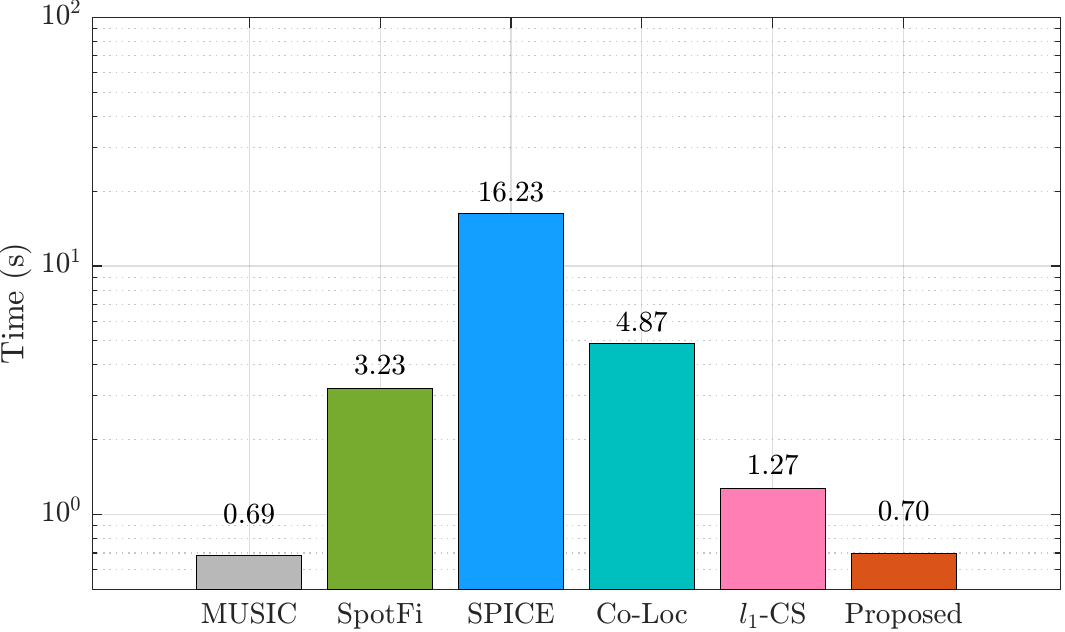}} 
    \caption{The time complexity comparison.} 
    \label{Time_Complexity}  
\end{figure}

We conduct an empirical analysis of the time complexity of Algorithm~\ref{algorithm_fusion} using the six angle estimation algorithms previously discussed. Here, the CSI data is processed using an Intel Core i7-12700K processor with 64 GB of RAM. The time required for MUSIC, SpotFi~\cite{kotaru2015spotfi}, SPICE~\cite{stoica2010spice}, Co-Loc~\cite{yang2023multiple}, $\ell_1$-norm CS~\cite{zhang20193d} and the proposed algorithm to perform localization is shown in Fig.~\ref{Time_Complexity}. The subspace methods, which require less traversal, generally exhibit lower overall computational time. However, SpotFi, in comparison to the other two subspace techniques, incurs additional processing time due to the subcarriers smoothing operation. Moreover, the SPICE and Co-Loc methods are performed with the baseline geometric method for timing. These optimization-based iterative methods are more time-consuming when dealing with a large number of parameters. In comparison to subspace methods, they retains a greater potential for optimization. The proposed DPD2URA algorithm requires approximately 0.7~s for execution. Specifically, I-SSMUSIC algorithm takes an average of 0.033 s, with each subspace-based fusion search averaging 0.166 s. The default number of search iterations is within four. 

\section{Conclusion and Future Work} \label{Conclusion} 
This work introduced WiCAL, the first practical 3D localization Wi-Fi system deploying multiple collaborative antenna arrays. The key of WiCAL lies in its innovative hardware solution for obtaining 3D AoA information through switched URAs. This solution ensures a smooth evolution of Wi-Fi standards, requiring only minimal adjustments to the physical layer. Additionally, an improved spatial smoothing MUSIC algorithm and a closest geometric point estimation were combined in this scheme. A DPD method was applied via subspace-based data fusion, enhanced by a progressive local traversal strategy. Experimental results have shown that WiCAL achieves a 15.6~cm median error in 3D localization. 

Future Wi-Fi standards, such as Wi-Fi 8 and beyond, are expected to leverage higher frequencies to substantially improve localization accuracy. At these frequencies, reduced environmental interference with the channel will contribute to enhanced precision~\cite{pegoraro2023rapid}. Furthermore, the expansion of RF chains and collaboration among multiple APs in future Wi-Fi networks promise even more accurate 3D positioning. Our study lays the foundation for multi-AP joint 3D positioning, data fusion, and array synchronization, marking a significant step forward in Wi-Fi-based 3D localization.

\bibliographystyle{IEEEtran}
\bibliography{Reference}

\begin{thebibliography}{10}
\providecommand{\url}[1]{#1}
\csname url@samestyle\endcsname
\providecommand{\newblock}{\relax}
\providecommand{\bibinfo}[2]{#2}
\providecommand{\BIBentrySTDinterwordspacing}{\spaceskip=0pt\relax}
\providecommand{\BIBentryALTinterwordstretchfactor}{4}
\providecommand{\BIBentryALTinterwordspacing}{\spaceskip=\fontdimen2\font plus
\BIBentryALTinterwordstretchfactor\fontdimen3\font minus \fontdimen4\font\relax}
\providecommand{\BIBforeignlanguage}[2]{{%
\expandafter\ifx\csname l@#1\endcsname\relax
\typeout{** WARNING: IEEEtran.bst: No hyphenation pattern has been}%
\typeout{** loaded for the language `#1'. Using the pattern for}%
\typeout{** the default language instead.}%
\else
\language=\csname l@#1\endcsname
\fi
#2}}
\providecommand{\BIBdecl}{\relax}
\BIBdecl

\bibitem{zhang2019efficient}
Y.~Zhang and K.~Psounis, ``Efficient indoor localization via switched-beam antennas,'' \emph{IEEE. Trans. Mob. Comput.}, vol.~19, no.~9, pp. 2101--2115, Jun. 2019.

\bibitem{10592075}
Y.~Yang \emph{et~al.}, ``Positioning using wireless networks: Applications, recent progress, and future challenges,'' \emph{IEEE J. Sel. Areas Commun.}, vol.~42, no.~9, pp. 2149--2178, Jul. 2024.

\bibitem{geraci2022integrating}
G.~Geraci, D.~L{\'o}pez-P{\'e}rez, M.~Benzaghta, and S.~Chatzinotas, ``Integrating terrestrial and non-terrestrial networks: {{3D}} opportunities and challenges,'' \emph{IEEE Commun. Mag.}, vol.~61, no.~4, pp. 42--48, Dec. 2022.

\bibitem{10608169}
B.~Zheng and F.~Liu, ``Random signal design for joint communication and {SAR} imaging towards low-altitude economy,'' \emph{IEEE Wirel. Commun. Lett.}, vol.~13, no.~10, pp. 2662--2666, Jul. 2024.

\bibitem{ayyalasomayajula2020deep}
R.~Ayyalasomayajula, A.~Arun, C.~Wu, S.~Sharma, A.~R. Sethi, D.~Vasisht, and D.~Bharadia, ``Deep learning based wireless localization for indoor navigation,'' in \emph{Proc. Annu. Int. Conf. Mob. Comput. Netw.(MobiCom)}, Apr. 2020, pp. 1--14.

\bibitem{arun2024wais}
A.~Arun, W.~Hunter, R.~Ayyalasomayajula, and D.~Bharadia, ``{WAIS}: Leveraging {{Wi-Fi}} for resource-efficient {SLAM},'' in \emph{Proc. Annu. Int. Conf. Mobile Syst. Appl. Service.(MobiSys)}, Jun. 2024, pp. 561--574.

\bibitem{tedeschini2024real}
B.~C. Tedeschini, G.~Kwon, M.~Nicoli, and M.~Z. Win, ``Real-time bayesian neural networks for {6G} cooperative positioning and tracking,'' \emph{IEEE J. Sel. Areas Commun.}, vol.~42, no.~9, pp. 2322--2338, Aug. 2024.

\bibitem{xu2024mimo}
C.~Xu and S.~Zhang, ``{{MIMO}} integrated sensing and communication exploiting prior information,'' \emph{IEEE J. Sel. Areas Commun.}, vol.~42, no.~9, pp. 2306--2321, Jul. 2024.

\bibitem{kotaru2017position}
M.~Kotaru and S.~Katti, ``Position tracking for virtual reality using commodity {{Wi-Fi}},'' in \emph{Proc. IEEE Conf. Comput. Vis. Pattern Recognit.}, Jul. 2017, pp. 68--78.

\bibitem{10599121}
S.~Kim \emph{et~al.}, ``Role of sensing and computer vision in {6G} wireless communications,'' \emph{IEEE Wirel. Commun.}, vol.~31, no.~5, pp. 264--271, Jul. 2024.

\bibitem{xiong2013arraytrack}
J.~Xiong and K.~Jamieson, ``{ArrayTrack}: {A} fine-grained indoor location system,'' in \emph{Proc. 10th USENIX Symp. Networked Syst. Des. Implement. (NSDI 13)}, Apr. 2013, pp. 71--84.

\bibitem{wang2013spectral}
D.~Wang, J.~Wang, X.~You, Y.~Wang, M.~Chen, and X.~Hou, ``Spectral efficiency of distributed {MIMO} systems,'' \emph{IEEE J. Sel. Areas Commun.}, vol.~31, no.~10, pp. 2112--2127, Oct. 2013.

\bibitem{gjengset2014phaser}
J.~Gjengset, J.~Xiong, G.~McPhillips, and K.~Jamieson, ``Phaser: {Enabling} phased array signal processing on commodity {{Wi-Fi}} access points,'' in \emph{Proc. Annu. Int. Conf. Mob. Comput. Netw.(MobiCom)}, Sep. 2014, pp. 153--164.

\bibitem{xie2018swan}
Y.~Xie, Y.~Zhang, J.~C. Liando, and M.~Li, ``{SWAN}: {Stitched} {{Wi-Fi}} antennas,'' in \emph{Proc. Annu. Int. Conf. Mob. Comput. Netw.(MobiCom)}, Oct. 2018, pp. 51--66.

\bibitem{gu2021tyrloc}
Z.~Gu, T.~He, J.~Yin, Y.~Xu, and J.~Wu, ``Tyrloc: A low-cost multi-technology {MIMO} localization system with a single {RF} chain,'' in \emph{Proc. Annu. Int. Conf. Mobile Syst. Appl. Service.(MobiSys)}, Jun. 2021, pp. 228--240.

\bibitem{an2020general}
Z.~An, Q.~Lin, P.~Li, and L.~Yang, ``General-purpose deep tracking platform across protocols for the internet of things,'' in \emph{Proc. Annu. Int. Conf. Mobile Syst. Appl. Service.(MobiSys)}, Jun. 2020, pp. 94--106.

\bibitem{kotaru2015spotfi}
M.~Kotaru, K.~Joshi, D.~Bharadia, and S.~Katti, ``Spotfi: Decimeter level localization using {Wi-Fi},'' in \emph{Proc. ACM SIGCOMM}, Aug. 2015, pp. 269--282.

\bibitem{he2020multi}
D.~He, X.~Chen, L.~Pei, F.~Zhu, L.~Jiang, and W.~Yu, ``{Multi-BS spatial spectrum fusion for 2-D DOA estimation and localization using UCA in massive {MIMO} system},'' \emph{IEEE Commun. Mag.}, vol.~70, pp. 1--13, Oct. 2020.

\bibitem{song2022rf}
R.~Song \emph{et~al.}, ``{RF-URL}: Unsupervised representation learning for {RF} sensing,'' in \emph{Proc. Annu. Int. Conf. Mob. Comput. Netw.(MobiCom)}, Oct. 2022, pp. 282--295.

\bibitem{soltanaghaei2018multipath}
E.~Soltanaghaei, A.~Kalyanaraman, and K.~Whitehouse, ``Multipath triangulation: Decimeter-level {Wi-Fi} localization and orientation with a single unaided receiver,'' in \emph{Proc. Annu. Int. Conf. Mobile Syst. Appl. Service.(MobiSys)}, Jun. 2018, pp. 376--388.

\bibitem{qian2018widar2}
K.~Qian, C.~Wu, Y.~Zhang, G.~Zhang, Z.~Yang, and Y.~Liu, ``Widar2.0: Passive human tracking with a single {Wi-Fi} link,'' in \emph{Proc. Annu. Int. Conf. Mobile Syst. Appl. Service.(MobiSys)}, Jun. 2018, pp. 350--361.

\bibitem{chen2023cross}
C.~Chen, G.~Zhou, and Y.~Lin, ``Cross-domain {Wi-Fi} sensing with channel state information: A survey,'' \emph{ACM Comput. Surv.}, vol.~55, no.~11, pp. 1--37, Feb. 2023.

\bibitem{sesyuk2022survey}
A.~Sesyuk, S.~Ioannou, and M.~Raspopoulos, ``A survey of {3D} indoor localization systems and technologies,'' \emph{Sensors}, vol.~22, no.~23, p. 9380, Dec. 2022.

\bibitem{farahsari2022survey}
P.~S. Farahsari, A.~Farahzadi, J.~Rezazadeh, and A.~Bagheri, ``A survey on indoor positioning systems for {IoT-based} applications,'' \emph{IEEE Internet Things J.}, vol.~9, no.~10, pp. 7680--7699, Feb. 2022.

\bibitem{liu2013accurate}
H.~Liu, J.~Yang, S.~Sidhom, Y.~Wang, Y.~Chen, and F.~Ye, ``Accurate {Wi-Fi} based localization for smartphones using peer assistance,'' \emph{IEEE. Trans. Mob. Comput.}, vol.~13, no.~10, pp. 2199--2214, Oct. 2013.

\bibitem{xiong2024fair}
R.~Xiong, K.~Yin, T.~Mi, J.~Lu, K.~Wan, and R.~C. Qiu, ``Fair beam allocations through reconfigurable intelligent surfaces,'' \emph{IEEE J. Sel. Areas Commun.}, vol.~42, no.~11, pp. 3095--3109, Jul. 2024.

\bibitem{tai2019toward}
T.-C. Tai, K.~C.-J. Lin, and Y.-C. Tseng, ``Toward reliable localization by unequal {AoA} tracking,'' in \emph{Proc. Annu. Int. Conf. Mobile Syst. Appl. Service.(MobiSys)}, Jun. 2019, pp. 444--456.

\bibitem{zhao2023nerf2}
X.~Zhao, Z.~An, Q.~Pan, and L.~Yang, ``Nerf2: Neural radio-frequency radiance fields,'' in \emph{Proc. Annu. Int. Conf. Mob. Comput. Netw.(MobiCom)}, Oct. 2023, pp. 1--15.

\bibitem{zhang20193d}
L.~Zhang and H.~Wang, ``{3D}-{WiFi}: {3D} localization with commodity{Wi-Fi},'' \emph{IEEE Sensors Journal}, vol.~19, no.~13, pp. 5141--5152, Feb. 2019.

\bibitem{qian2017enabling}
K.~Qian, C.~Wu, Z.~Yang, Z.~Zhou, X.~Wang, and Y.~Liu, ``Enabling phased array signal processing for mobile {Wi-Fi} devices,'' \emph{IEEE. Trans. Mob. Comput.}, vol.~17, no.~8, pp. 1820--1833, Nov. 2017.

\bibitem{wu2021witraj}
D.~Wu \emph{et~al.}, ``{WiTraj}: Robust indoor motion tracking with {Wi-Fi} signals,'' \emph{IEEE. Trans. Mob. Comput.}, vol.~22, no.~5, pp. 3062--3078, Dec. 2021.

\bibitem{stoica2005spectral}
P.~Stoica \emph{et~al.}, \emph{Spectral analysis of signals}.\hskip 1em plus 0.5em minus 0.4em\relax Pearson Prentice Hall Upper Saddle River, NJ, 2005.

\bibitem{nanzer2021distributed}
J.~A. Nanzer, S.~R. Mghabghab, S.~M. Ellison, and A.~Schlegel, ``Distributed phased arrays: Challenges and recent advances,'' \emph{IEEE Trans. Microw. Theory Tech.}, vol.~69, no.~11, pp. 4893--4907, Jul. 2021.

\bibitem{yang2023multiple}
S.~Yang, D.~Zhang, R.~Song, P.~Yin, and Y.~Chen, ``Multiple {Wi-Fi} access points co-localization through joint {AoA} estimation,'' \emph{IEEE. Trans. Mob. Comput.}, vol.~23, no.~2, pp. 1488--1502, Jan. 2023.

\bibitem{edition2009antennas}
L.~V. Blake and M.~W. Long, \emph{Antennas: Fundamentals, design, measurement}.\hskip 1em plus 0.5em minus 0.4em\relax SciTech Publishing, Raleigh, 2009.

\bibitem{stutzman2012antenna}
W.~L. Stutzman and G.~A. Thiele, \emph{Antenna theory and design}.\hskip 1em plus 0.5em minus 0.4em\relax John Wiley \& Sons, 2012.

\bibitem{van2002optimum}
H.~L. Van~Trees, \emph{Optimum array processing: {Part} {IV} of detection, estimation, and modulation theory}.\hskip 1em plus 0.5em minus 0.4em\relax John Wiley \& Sons, 2002.

\bibitem{tirer2015high}
T.~Tirer and A.~J. Weiss, ``High resolution direct position determination of radio frequency sources,'' \emph{IEEE Signal Process. Lett.}, vol.~23, no.~2, pp. 192--196, Nov. 2015.

\bibitem{wang2022computationally}
Z.~Wang, K.~Hao, Y.~Sun, L.~Xie, and Q.~Wan, ``A computationally efficient direct position determination algorithm based on {OFDM} system,'' \emph{IEEE Commun. Lett.}, vol.~27, no.~3, pp. 841--845, Dec. 2022.

\bibitem{Halperin_csitool}
D.~Halperin, W.~Hu, A.~Sheth, and D.~Wetherall, ``Tool release: Gathering 802.11n traces with channel state information,'' \emph{ACM SIGCOMM Comput. Commun. Rev.}, vol.~41, no.~1, p.~53, Jan. 2011.

\bibitem{shan1985spatial}
T.-J. Shan, M.~Wax, and T.~Kailath, ``On spatial smoothing for direction-of-arrival estimation of coherent signals,'' \emph{IEEE Trans. Acoust. Speech Signal Process.}, vol.~33, no.~4, pp. 806--811, Aug. 1985.

\bibitem{stoica2010spice}
P.~Stoica, P.~Babu, and J.~Li, ``{SPICE}: {A} sparse covariance-based estimation method for array processing,'' \emph{IEEE Trans. Signal Process.}, vol.~59, no.~2, pp. 629--638, Nov. 2010.

\bibitem{li2023riscan}
C.~Li~et al., ``{RIScan}: {RIS}-aided multi-user indoor localization using cots {Wi-Fi},'' in \emph{Proc. ACM SenSys}, Apr. 2023, pp. 445--458.

\bibitem{pegoraro2023rapid}
J.~Pegoraro, J.~O. Lacruz, F.~Meneghello, E.~Bashirov, M.~Rossi, and J.~Widmer, ``{RAPID}: Retrofitting {IEEE} 802.11ay access points for indoor human detection and sensing,'' \emph{IEEE. Trans. Mob. Comput.}, vol.~23, no.~5, pp. 4501--4519, Jul. 2023.

\end{thebibliography}

\begin{IEEEbiography}[{\includegraphics[width=1in,height=1.25in,clip,keepaspectratio]{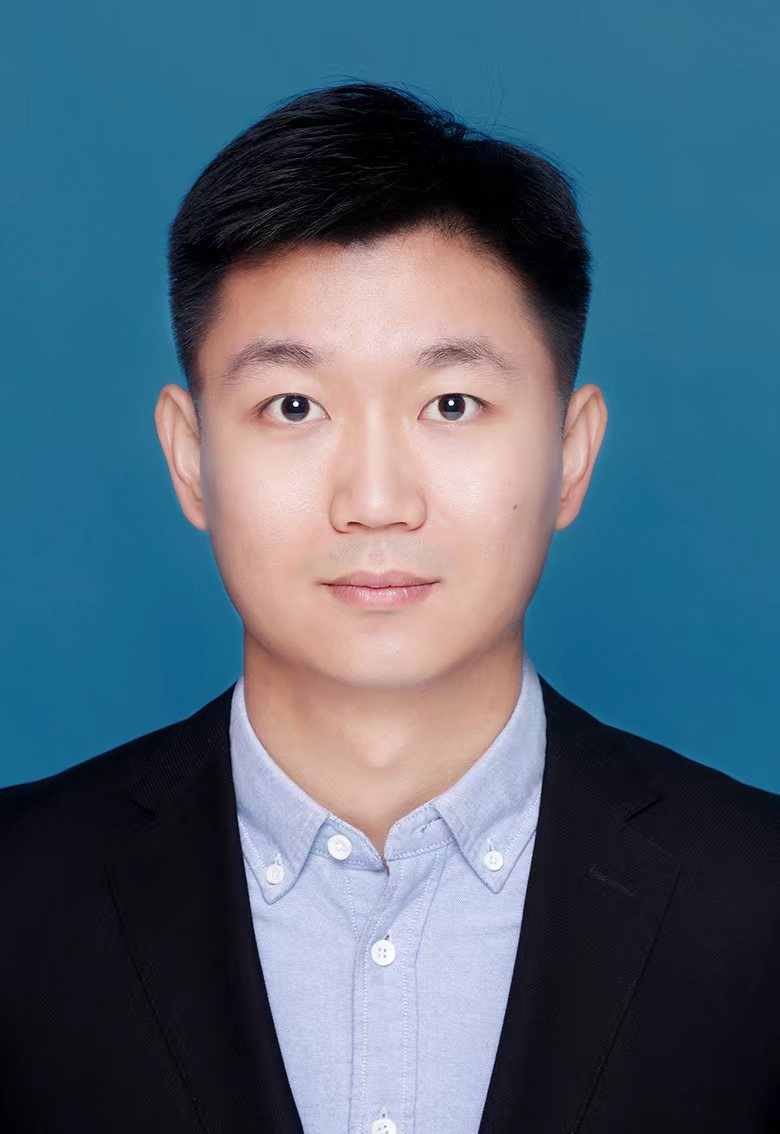}}]{Fuhai Wang}
(Graduate Student Member, IEEE) received the B.S. degree in communication engineering from Chongqing University of Posts and Telecommunications, China, in 2018, and the M.S. degree in materials science from University of Chinese Academy of Sciences, China, in 2021. He is currently pursuing the Ph.D. degree with the Institute of Artificial Intelligence, Huazhong University of Science and Technology, China. His current research interests include wireless localization and sensing, metamaterials imaging, and RF-based scene reconstruction. \end{IEEEbiography}

\begin{IEEEbiography}[{\includegraphics[width=1in,height=1.25in,clip,keepaspectratio]{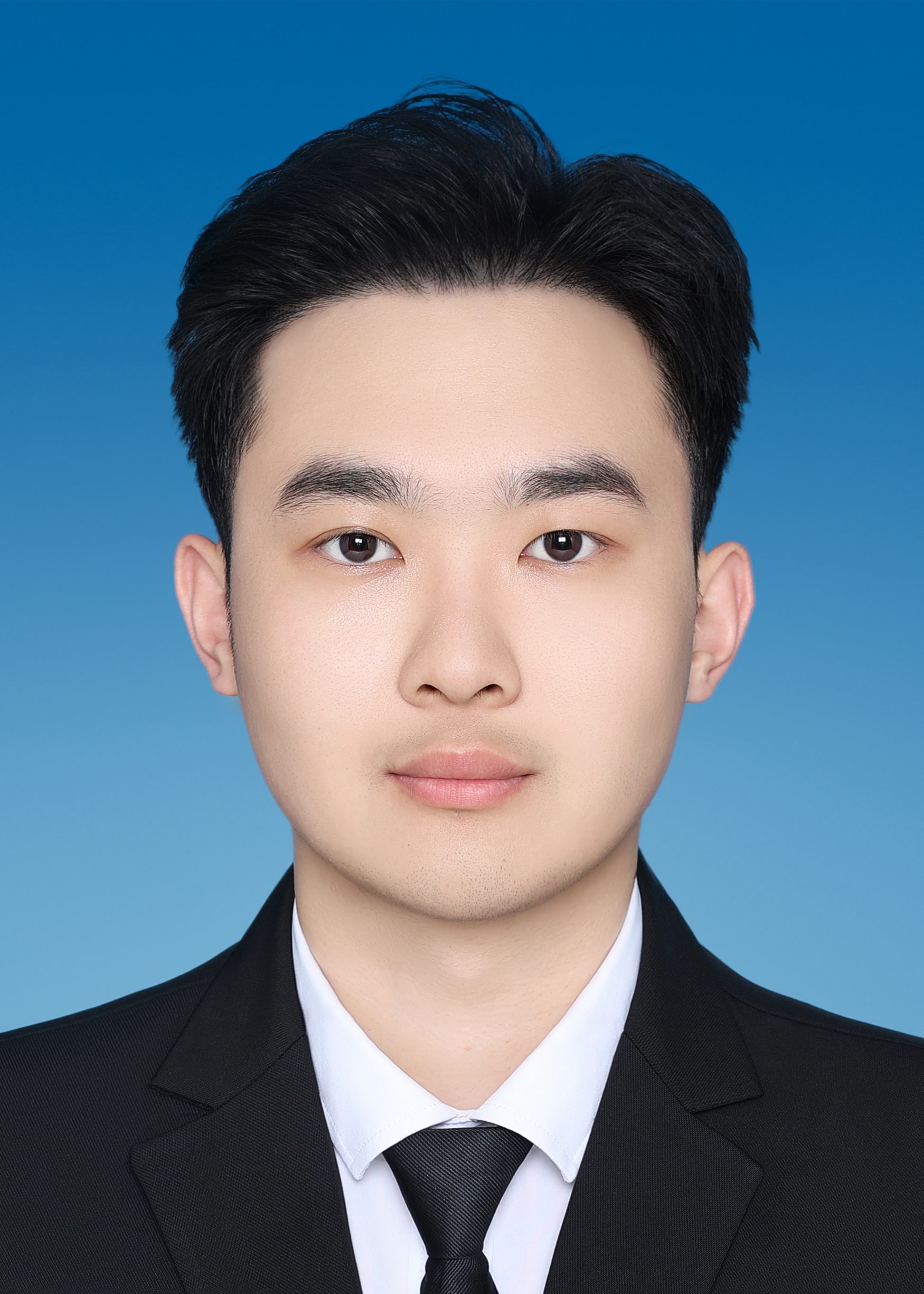}}]{Zhe Li}
received the B.S. degree in electronic information and communications from the School of Electronics and Information Engineering at Harbin Institute of Technology, China, in 2022. He is currently pursuing the M.S. degree with the School of Electronic Information and Communications, Huazhong University of Science and Technology, China. His current research areas include WiFi indoor positioning, wireless sensing, and radar signal processing.
\end{IEEEbiography}

\begin{IEEEbiography}[{\includegraphics[width=1in,height=1.25in,clip,keepaspectratio]{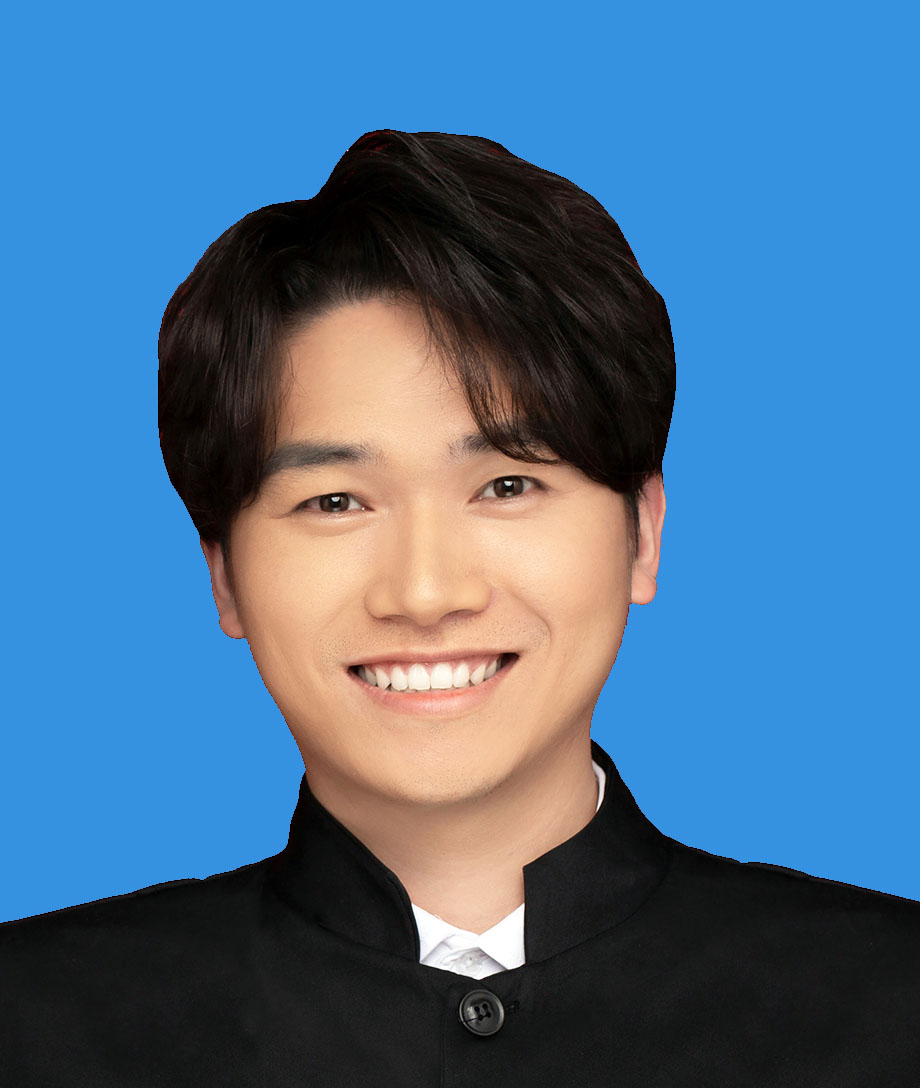}}]{Rujing Xiong}
(Graduate Student Member, IEEE) received the B.S. degree in bioinformatics from Zhengzhou University, China, in 2017, and the M.S. degree in electronics and communication engineering from Central South University, China, in 2020. He is currently pursuing the Ph.D. degree with the School of Electronic Information and Communications, Huazhong Univerisity of Science and Technology, China. His research interests include wireless communication, non-convex optimization, extremely large antenna arrays and reconfigurable intelligent surfaces.\end{IEEEbiography}

\begin{IEEEbiography}[{\includegraphics[width=1in,height=1.25in,clip,keepaspectratio]{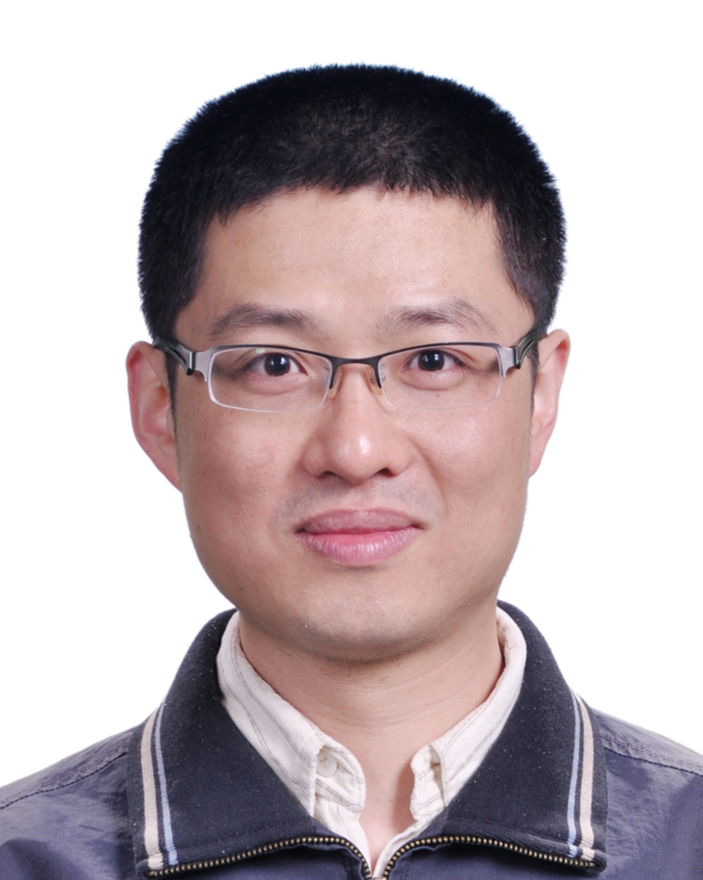}}]{Tiebin Mi}
(Member, IEEE) received the B.E. degree in computer science from Xidian University, China, in 2002, and the Ph.D. degree in electrical engineering from the Institute of Acoustics, Chinese Academy of Sciences, China, in 2010. Currently, he is a Lecturer (an Assistant Professor) with the School of Electronic Information and Communications, Huazhong University of Science and Technology, China. His current research interests include wireless communications, high-dimensional signal processing, random matrix theory, and reconfigurable intelligent surfaces.\end{IEEEbiography}

\begin{IEEEbiography}[{\includegraphics[width=1in,height=1.25in,clip,keepaspectratio]{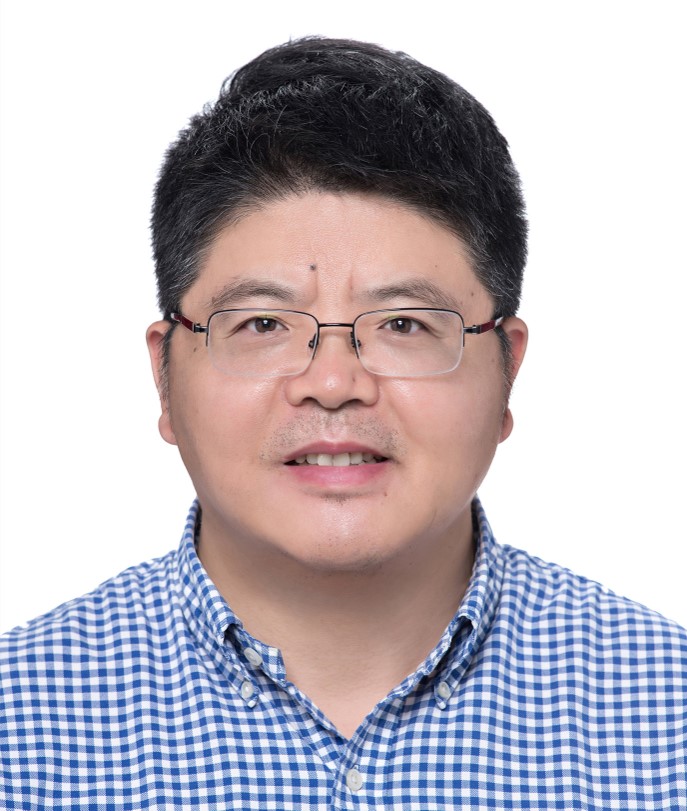}}]{Robert Caiming Qiu}
(Fellow, IEEE) received the Ph.D. degree in electrical engineering from New York University (former Polytechnic University) Brooklyn, NY, USA. He joined the School of Electronic Information and Communications, Huazhong University of Science and Technology, China, as a Full Professor, in 2020. Before joining HUST, he was an Associate Professor with the Department of Electrical and Computer Engineering, Center for Manufacturing Research, Tennessee Tech University, Cookeville, TN, USA, in 2003, where he became a Professor in 2008. He founded and served as the CEO and President of Wiscom Technologies Inc., a company specializing in manufacturing and marketing WCDMA chipsets. In 2003, he was acquired by Intel. He was with GTE Laboratories Inc. (now Verizon), Waltham, MA, USA, and Bell Laboratories, Lucent, Whippany, NJ, USA. He has coauthored Cognitive Radio Communication and Networking: Principles and Practice (John Wiley, 2012) and Cognitive Networked Sensing: A Big Data Way (Springer, 2013) and authored Big Data and Smart Grid (John Wiley, 2015). He has authored over 100 journal articles/book chapters and 120 conference papers. He was a Guest Book Editor of Ultra-Wideband (UWB) Wireless Communications (New York: Wiley, 2005) and three Special Issues on UWB, including the IEEE Journal on Selected Areas in Communications, IEEE Transactions on Vehicular Technology, and IEEE Transactions on Smart Grid. Furthermore, he has made 15 contributions to 3GPP and IEEE standards bodies. He has served as a TPC Member for GLOBECOM, ICC, WCNC, and MILCOM. He has also served as an Associate Editor for IEEE Transactions on Vehicular Technology and other international journals. \end{IEEEbiography}

\end{document}